\documentclass[]{interact}

\usepackage{epstopdf}
\usepackage[caption=false]{subfig}

\usepackage{amsmath}
\usepackage{amsthm}
\usepackage{amssymb}
\usepackage{ulem}
\newcommand{\pkg}[1]{{\fontseries{b}\selectfont #1}}

\let\proglang=\textsf

\newcommand{\bftheta}{{\boldsymbol \theta}}
\newcommand{\bfbeta}{{\boldsymbol \beta}}
\newcommand{\bfSig}{{\bf \Sigma}}
\newcommand{\bfs}{{\bf s}}
\newcommand{\bfx}{{\bf x}}

\usepackage[numbers,sort&compress]{natbib}
\bibpunct[, ]{[}{]}{,}{n}{,}{,}
\makeatletter
\def\NAT@def@citea{\def\@citea{\NAT@separator}}
\makeatother

\theoremstyle{plain}

\theoremstyle{definition}

\theoremstyle{remark}

\begin{document}


\title{Bayesian inference for high-dimensional nonstationary Gaussian processes}

\author{
\name{Mark D. Risser\textsuperscript{a}\thanks{Corresponding author: Mark D. Risser. Email: {\tt mdrisser@lbl.gov}} and Daniel Turek\textsuperscript{b}}
\affil{\textsuperscript{a}Lawrence Berkeley National Laboratory, Berkeley, CA, US; \textsuperscript{b}Williams College, Williamstown, MA, US}
}

\maketitle

\begin{abstract}
In spite of the diverse literature on nonstationary spatial modeling and approximate Gaussian process (GP) methods, there are no general approaches for conducting fully Bayesian inference for moderately sized nonstationary spatial data sets on a personal laptop. For statisticians and data scientists who wish to learn about spatially-referenced data and conduct posterior inference and prediction with appropriate uncertainty quantification, the lack of such approaches and corresponding software is a significant limitation. In this paper, we develop methodology for implementing formal Bayesian inference for a general class of nonstationary GPs. Our novel approach uses pre-existing frameworks for characterizing nonstationarity in a new way that is applicable for small to moderately sized data sets via modern GP likelihood approximations. Posterior sampling is implemented using flexible MCMC methods, with nonstationary posterior prediction conducted as a post-processing step. We demonstrate our novel methods on two data sets, ranging from several hundred to several thousand locations, and compare our methodology with related statistical methods that provide off-the-shelf software. All of our methods are implemented in the freely available \pkg{BayesNSGP} software package for \proglang{R}.
\end{abstract}


\section{Introduction} \label{section1}
Gaussian processes (GPs) are an extremely popular tool in modern statistical modeling, with broad application in spatial and environmental statistics as well as machine learning and emulation of complex mathematical and physical models. However, methods and software for conducting formal Bayesian analysis for general classes of nonstationary GPs (wherein the spatial dependence structure varies over the domain of interest; see \cite{Risser2016} for a review of recent methods) do not currently exist, which presents a problem for spatial statisticians and data scientists who either (a) require tools for posterior inference and prediction for real-world spatial data sets or (b) are developing new nonstationary methods and need to compare against existing approaches. The broad literature on nonstationary methods are generally difficult to implement because most are highly parameterized and require highly specialized algorithms for model fitting \citep[e.g.,][]{Sampson1992, Higdon1998, damian2001, Schmidt2003, Paciorek2006}. Furthermore, no existing nonstationary methods have been extended to include modern advances in approximate Gaussian process modeling (see \cite{heaton2018case} for a summary of recent methods) which enable inference for large data sets (other than the methods based on Gaussian Markov random fields, e.g., \cite{Lindgren2011}).

In light of these shortcomings, we develop a novel and highly flexible framework for conducting fully Bayesian inference for nonstationary spatial GPs, with a particular focus on implementation and enabling a general statistician to analyze moderately large real-world data sets. At the core of our methodology is a convolution-based nonstationary covariance function that combines existing frameworks for characterizing nostationarity \citep[e.g.,][]{Higdon1998,Paciorek2006,Risser2015} in a novel way: spatially-varying parameters can be specified either deterministically (using covariates or basis functions) or stochastically (using approximate Gaussian processes), with stationarity as a special case. Furthermore, our methods are scalable to high dimensional data sets via a framework for embedding the covariance function into one of two approximate GP methods. Posterior prediction for the GPs at unobserved locations can be conducted for both exact and approximate GP methods as a  post-processing step. And, critically, we provide the \pkg{BayesNSGP} package for \proglang{R} that for the first time enables off-the-shelf functionality for fully Bayesian, convolution-based nonstationary GP modeling of  moderately large data sets (up to at least 50,000 measurement locations; see \cite{risser2020nonstationary}). The software package relies on underlying tools from the \pkg{nimble} \citep{nimble_jcgs} package to implement flexible Markov chain Monte Carlo methods for sampling the highly correlated parameter spaces common to nonstationary Gaussian process models. All of our methods are implemented for use on a personal laptop and do not require a custom computing environment, making them accessible to the entire statistics community.

It should be noted that there are a variety of software tools available for analyzing both nonstationary spatial data and very large data sets. For example, \proglang{R} packages like \pkg{fields} \citep{Nychka2014} and \pkg{convoSPAT} \citep{Risser2017} provide software for Frequentist inference, while \pkg{INLA} \citep{Lindgren2011, Ingebrigtsen2013, Fuglstad2015, Lindgren2015}, \pkg{spBayes} \citep{Finley2007}, and \pkg{tgp} \citep{Gramacy2007}  provide software for Bayesian analysis of nonstationary spatial data. More recently, the \pkg{spNNGP}  \citep{finley2017spnngp,finley2020r} and \pkg{GPvecchia} \citep{GPvecchia} packages provide a variety of spatial regression tools (albeit not nonstationary, and \pkg{GPvecchia} is not Bayesian) for very large data sets. Outside of \proglang{R}, there are open source tools on Github that implement, e.g., the multi-resolution approximation  \citep[MRA;][]{katzfuss2017multi} for very large to massive data sets \citep{huang2019pushing,blake2019a}. While these are a flexible set of tools, none of these approaches implement fully Bayesian analysis for general classes of convolution-based methods \citep[e.g.,][]{Higdon1999, Paciorek2006}, which both flexibly model nonstationary processes (by allowing the parameters of the covariance function to vary over space) and also yield interpretable summaries of how and why a process exhibits nonstationarity \citep[see][]{Risser2015, Risser2017}. Furthermore, most of the \proglang{R} packages cannot appropriately model even moderately large spatial data sets. While in principle the MRA software \citep{huang2019pushing,blake2019a}, \pkg{spNNGP} package \citep{finley2017spnngp,finley2020r}, and and \pkg{GPvecchia} package \citep{GPvecchia} could be either fully Bayesian and/or incorporate a nonstationary covariance function, the corresponding implementation is nontrivial and in any case has not been done to date. And, it should be noted that (at least) the MRA software is tackling a very different problem, namely fast analysis of massive data sets with custom computing environments, whereas we seek to provide analysis tools for use on a personal laptop.

As a final note, we mention that our methods assume a Gaussian likelihood for the data, making it inappropriate for data analyses that require non-Gaussian likelihoods (e.g., count data or highly skewed data). Of course, our methodology could be applied in a hierarchical model that assigns a GP to a latent process in the statistical model (e.g., the log intensity function for Poisson data); in this case, the latent GP would undergo MCMC sampling, which is nontrivial when dealing with small data sets. We refer the interested reader to \cite{bradley2017bayesian} and \cite{zilber2019vecchia} for recent developments in modeling large dependent data for more general likelihoods from the natural exponential family (as well as \cite{finley2020r}, which incorporates non-Gaussian likelihoods).

The paper proceeds as follows. Section \ref{CGPM} outlines a canonical Bayesian nonstationary Gaussian process model as well as prediction for unobserved locations; Section \ref{section3} outlines our novel modeling framework for the nonstationary covariance function. Section \ref{section4} describes our approach to approximate Gaussian process inference for large data sets and corresponding approximate posterior prediction for unobserved locations. Section \ref{implem} briefly describes software for implementing our novel statistical methodology, and Section \ref{section6} illustrates the usefulness of our methods and makes comparison with existing methods through two examples. Section \ref{sec_discussion} concludes the paper.

\section{Canonical Bayesian Gaussian process model} \label{CGPM}

Let $\{ z(\bfs) : \bfs \in G\}$ be the observed value of a univariate spatial process over a domain $G \subset \mathcal{R}^d$, with $d\geq 1$. A general framework for modeling $z(\bfs)$ as a spatial Gaussian process can be defined via the linear mixed model 
\begin{equation} \label{CANONmodel}
z({\bf s}) = y({\bf s}) + \varepsilon({\bf s}), 
\end{equation}
where $E[z({\bf s})] = y({\bf s})$, $y(\cdot)$ is a spatial random effect, and $\varepsilon(\cdot)$ is a stochastic component that represents measurement error or microscale variability and is independently distributed as ${N}(0, \tau^2(\bfs))$ such that $\varepsilon(\cdot)$ and $y(\cdot)$ are independent. The spatial random effect is modeled as a parametric Gaussian process, denoted $y(\cdot) \sim GP\big( {\bf x}(\cdot)^\top \bfbeta, C_y(\cdot,\cdot; \bftheta_y)\big)$, such that $E[y(\bfs)] = {\bf x}(\bfs)^\top \bfbeta$ is a 
linear (deterministic) mean function in a set of $p-1$ covariates (with an intercept, i.e., $\bfx(\bfs) = \big(1, x_1(\bfs), \dots, x_{p-1}(\bfs)\big)^\top \in \mathcal{R}^{p}$). The covariance function $C_y$ is assumed known up to a vector of parameters $\bftheta_y$ and describes the covariance between the process $y(\cdot)$ as  
\[
C_y(\bfs, \bfs'; \bftheta_y) \equiv Cov\big(y(\bfs), y(\bfs') \big),
\]
for all $\bfs, \bfs' \in G$. Finally, we suppose that the error variance process $\tau^2(\cdot)$ is known up to a vector of parameters $\bftheta_z$.

For a fixed, finite set of $N$ observed spatial locations $\mathcal{S}_O = \{{\bf s}_1, ... , {\bf s}_N\}\in G$, (\ref{CANONmodel}) implies that the random (observed) vector ${\bf z}_O = \left[ z({\bf s}_1), ... , z({\bf s}_N) \right]^\top$ has a multivariate Gaussian distribution 
\begin{equation} \label{Zcond}
p({\bf z}_O | {\bf y}_O, \bftheta_z ) = {N}\big({\bf y}_O, {\bf \Delta}(\bftheta_z)\big),
\end{equation}
where ${\bf \Delta}(\bftheta_z) = diag[\tau^2(\bfs_1), \dots, \tau^2(\bfs_N)]$. Conditional on the other parameters in the model, the process vector ${\bf y}_O = \left[ y({\bf s}_1), ... , y({\bf s}_N) \right]^\top$ is distributed as 
\begin{equation} \label{Ycond}
p({\bf y}_O | \bfbeta, \bftheta_y) = {N}\big({\bf X}_O\bfbeta, {\bf \Omega}(\bftheta_y) \big),
\end{equation}
where ${\bf X}_O = [\bfx(\bfs_1)^\top, \dots, \bfx(\bfs_N)^\top]^\top$ and the elements of ${\bf \Omega}(\bftheta_y)$ are $\Omega_{ij} \equiv C_y({\bf s}_i, {\bf s}_j; \bftheta_y)$. Given the Gaussian distributions in (\ref{Zcond}) and (\ref{Ycond}), it is often useful to integrate over the process $y(\cdot)$ to arrive at the marginal distribution for $z(\cdot)$, which is
\begin{equation} \label{Zmarg}
p({\bf z}_O | \bfbeta, \bftheta) = \int p({\bf z}_O | {\bf y}_O, \bftheta_z ) p({\bf y}_O | \bfbeta, \bftheta_y) d{\bf y}_O = {N}\big({\bf X}_O\bfbeta, {\bf \Delta}(\bftheta_z) + {\bf \Omega}(\bftheta_y) \big),
\end{equation}
where $\bftheta = (\bftheta_z, \bftheta_y)$. The covariance function for the marginalized process is
\begin{equation} \label{Zcov}
C_z(\bfs, \bfs'; \bftheta) = C_y(\bfs, \bfs'; \bftheta_y) + \tau(\bfs)\tau(\bfs')I_{\{\bfs = \bfs'\}}, \hskip4ex \text{for all } \bfs, \bfs' \in G,
\end{equation}
where $I_{\{\cdot\}}$ is an indicator function. 

To complete the Bayesian specification of this model, we define prior distributions for the unknown mean and covariance parameters $p(\bfbeta, \bftheta)$, where these priors are assumed to be independent (i.e., $p(\bfbeta, \bftheta) = p(\bfbeta) p(\bftheta)$) and noninformative (see the Supplemental Materials for more details on the priors used in our implementation). All inference for $\bfbeta$ and $\bftheta$ is based on the marginalized posterior for these parameters conditional on ${\bf z}_O$:
\begin{equation} \label{posteriorZ}
p(\bfbeta, \bftheta | {\bf z}_O) \propto p({\bf z}_O | \bfbeta, \bftheta) p(\bfbeta) p(\bftheta).
\end{equation}
Regardless of the form of the priors on $\bfbeta$ and $\bftheta$, the posterior distribution (\ref{posteriorZ}) is not available in closed form, and so we must resort to Markov chain Monte Carlo (MCMC) methods to conduct inference on $\bfbeta$ and $\bftheta$. Providing methods for sampling from this posterior for a variety of covariance models is a primary objective of this paper.

Posterior prediction of the process $y(\cdot)$ for either the observed locations $\mathcal{S}_O$ or a distinct set of $M$ unobserved locations $\mathcal{S}_P = \{{\bf s}^*_1, ... , {\bf s}^*_M\}\in G$ is straightforward given the Gaussian process assumptions used here. Define ${\bf y}_P= \left( y({\bf s}^*_1), ... , y({\bf s}^*_M) \right)^\top$ and ${\bf y} = ({\bf y}_O, {\bf y}_P)$; the predictive distribution of interest is then
\begin{equation} \label{ppd}
p({\bf y} | {\bf z}_O ) = \int_{\bfbeta, \bftheta} p({\bf y}, \bfbeta, \bftheta | {\bf z}_O) d\bfbeta d\bftheta = \int_{\bfbeta, \bftheta} p({\bf y}| \bfbeta, \bftheta, {\bf z}_O) p(\bfbeta, \bftheta | {\bf z}_O) d\bfbeta d\bftheta.
\end{equation}
Based on the Gaussian assuptions, $p({\bf y}| \bfbeta, \bftheta, {\bf z}_O)$ is $N\left( {\bf m}_{{\bf y}|{\bf z}_O}, {\bf C}_{{\bf y}|{\bf z}_O} \right)$, where
\[
{\bf m}_{{\bf y}|{\bf z}_O} = {\bf X}\bfbeta + {\bf C}_{{\bf y}, {\bf z}_O} {\bf C}_{{\bf z}_O}^{-1} ({\bf z}_O - {\bf X}_O\bfbeta); \hskip3ex {\bf C}_{{\bf y}|{\bf z}_O} = {\bf C}_{\bf y} - {\bf C}_{{\bf y}, {\bf z}_O} {\bf C}_{{\bf z}_O}^{-1} {\bf C}_{{\bf z}_O, {\bf y}}.
\]
(Here, ${\bf C}_{(\cdot)}$ is the covariance or cross-covariance matrix corresponding to $(\cdot)$.) The other component under the integral on the far right hand side of (\ref{ppd}) is the posterior (\ref{posteriorZ}); hence, in practice, given a set of posterior samples $\{ \bfbeta_l, \bftheta_l : l = 1, \dots, L \}$ (obtained via MCMC), a Monte Carlo estimate of (\ref{ppd}) is obtained via
\[
p({\bf y} | {\bf z}_O ) \approx \sum_{l=1}^L p\left(  {\bf y} | {\bf z}_O, \bfbeta_l, \bftheta_l \right).
\]

\section{Nonstationary covariance function modeling} \label{section3}

Among the diverse literature on approaches for modeling a nonstationary covariance function, one of the more intuitive and flexible methods involves allowing the parameters of the covariance function $C_y$ to vary over space, the so-called spatially-varying parameters approach. {This approach is based on a process convolution approach to spatial modeling \citep[see, e.g.,][]{Higdon1998}, wherein a general nonstationary spatial stochastic process $y(\cdot)$ on $G \subset \mathcal{R}^d$ can be defined via}
\begin{equation*} \label{kernelconvolution}
y({\bf s}) = \int_{G} K_{\bf s}({\bf u}) dW({\bf u})
\end{equation*}
{\citep{thiebaux76,thiebaux_pedder}, where $W(\cdot)$ is a $d$-dimensional stochastic process and $K_{\bf s}(\cdot)$ is a spatially-varying kernel function with finite first and second moments. If $W(\cdot)$ is chosen to be Gaussian white noise, the resulting covariance function is}
\begin{equation*} \label{Hig_NS}
C({\bf s}, {\bf s'}) = \int_{\mathcal{R}^d}  K_{\bf s}({\bf u}) K_{\bf s'}({\bf u}) d{\bf u},
\end{equation*}
{which is a nonstationary covariance function. If the kernel functions are furthermore specified to be $d$-variate Gaussian densities centered at $\bfs$ with covariance matrix $\bfSig(\bfs)$, the above integral can be calculated analytically \citep{Higdon99}; however, the resulting covariance function has the undesirable property of yielding process realizations that are infinitely differentiable. In their seminal paper, \cite{Paciorek2006} derive a generalization of the covariance function in \cite{Higdon99} based on the Mat\'ern correlation function, which can yield process realizations ranging from non-differentiable to infinitely differentiable. A variety of papers \citep[e.g.,][]{Risser2015} provide a slight generalization of \cite{Paciorek2006} to yield the covariance function}
\begin{equation} \label{PScov}
C_y(\bfs, \bfs'; \bftheta) = \sigma(\bfs) \sigma(\bfs') \frac{\left|\bfSig(\bfs)\right|^{1/4}\left|\bfSig(\bfs')\right|^{1/4}}{\left|\frac{\bfSig(\bfs') + \bfSig(\bfs')}{2} \right|^{1/2}} \mathcal{M}_\nu\left(\sqrt{Q(\bfs, \bfs')}\right), \hskip3ex \bfs, \bfs' \in G,
\end{equation}
where
\begin{equation} \label{Qij}
Q(\bfs, \bfs') = (\bfs - \bfs')^\top \left(\frac{\bfSig(\bfs) + \bfSig(\bfs')}{2}\right)^{-1}(\bfs - \bfs'),
\end{equation}
and $\mathcal{M}_\nu(\cdot)$ is the Mat\'ern correlation function with smoothness $\nu$ (note, however, that $C_y$ is non-negative definite for any valid correlation function over $\mathcal{R}^d$, $d \geq 1$). In (\ref{PScov}), $\sigma(\cdot)$ is a spatially-varying standard deviation process and $\bfSig(\cdot)$ is a spatially-varying anisotropy process that controls the range and direction of dependence. {The benefit of using (\ref{PScov}) is that one can specify the differentiability of the corresponding spatial process (via the smoothness parameter $\nu$) within a nonstationary covariance function.} 
{Furthermore, t}he covariance function defined via (\ref{Zcov}) and (\ref{PScov}) is highly flexible, as it defines parameter processes $\sigma(\cdot)$ and $\bfSig(\cdot)$---and $\tau(\cdot)$, the standard deviation process for the error $\varepsilon(\cdot)$, when considering $C_z$---over an infinite-dimensional space (i.e., $G \subset \mathcal{R}^d$).
{We refer the interested reader to \cite{Risser2015, Risser2017, Risser2019} for a variety of examples of the possible shapes and spatial correlation patterns that (\ref{Zcov}) can produce.}

In practice, {since the parameter processes are defined over an infinite-dimensional space, they must be parameterized in some way such that} implementation is feasible. We now outline a variety of approaches for parameterizing and regularizing these processes, all of which involve modeling the parameter processes as spatially-varying fields. The various combinations of these different approaches constitute a novel contribution to the literature on nonstationary covariance function modeling.

\subsection{Scalar standard deviation processes} \label{SDproc}

The processes $\tau(\cdot)$ and $\sigma(\cdot)$ represent standard deviations for the error $\varepsilon(\cdot)$ and process $y(\cdot)$, respectively, and by definition are strictly positive. A variety of models can be specified for scalar (i.e., univariate) spatial processes defined on the positive real line. Several are outlined below. 

\subsubsection{Spatial constants}

While the main point of the methods in this paper is to allow the variance/covariance properties to vary over space, for completeness we also define spatially-constant error and spatial standard deviations:
\begin{equation} \label{constSD}
\tau(\bfs) \equiv \delta, \hskip3ex \sigma(\bfs) \equiv \alpha \hskip3ex \text{for all } \bfs \in G.
\end{equation}

\subsubsection{Log-linear regression}

The simplest approach for modeling spatially-varying $\tau(\cdot)$ and $\sigma(\cdot)$ is a regression model that is linear in a set of spatial covariates on the log scale:
\begin{equation} \label{LLR}
\log \tau(\bfs) = \bfx_\tau(\bfs)^\top \boldsymbol{\delta}, \hskip3ex \log \sigma(\bfs) = \bfx_\sigma(\bfs)^\top \boldsymbol{\alpha}, 
\end{equation}
where $\bfx_{(\cdot)}(\bfs) \in \mathcal{R}^{p_{(\cdot)}}$ is a set of fully-observed covariates (including an intercept) and $\boldsymbol{\delta}$ and $\boldsymbol{\alpha}$ are vectors of regression coefficients. This is likely the most parsimonious representation of the process, as it involves only $p_{(\cdot)}$ parameters. Note that in this framework, $\bfx_{(\cdot)}(\bfs)$ could represent either physical covariates (for a spatial process, e.g., orography or land use categories) or basis functions (e.g., wavelets, splines, or principal components).  

\subsubsection{Approximation to a stationary Gaussian process} \label{approxGP}

\noindent While the regression framework is highly parsimonious, it specifies a rigid relationship between the processes $\tau(\cdot)$ and $\sigma(\cdot)$ and the choice of covariates. Furthermore, as is often a challenge in spatial regression, the covariate vector  must be fully observed over the spatial domain, including any prediction location of interest. An additional degree of flexibility can be gained by modeling the natural logarithm of either $\tau(\cdot)$ or $\sigma(\cdot)$ as themselves stationary Gaussian processes with Mat\'ern covariance. For simplicity of notation, in the remainder of this section we generically use $\phi$ to represent either standard deviation process. So, for $\phi \in \{\tau, \sigma\}$, we can model 
\begin{equation} \label{sgp_phi}
\log \phi(\bfs) \sim GP\big(\mu_\phi, C_\phi(\cdot; \rho_\phi, \sigma_\phi, \nu_\phi)\big),
\end{equation}
where $E[\log \phi(\bfs)] = \mu_\phi$ and $C_\phi$ is the stationary Mat\'ern covariance function with standard deviation $\sigma_\phi$, spatial range $\rho_\phi$, and smoothness $\nu_\phi$. \cite{paciorek2004nonstationary} use this statistical model for the anisotropy components (see Section \ref{Anis_proc}), but \cite{Paciorek2006} found the MCMC computations to be slow (likely due to the fact that the parameter space for the model in Eq.~\ref{sgp_phi} is of dimension $N+4$). As an alternative, \cite{Paciorek2006} suggest using a basis function approximation to a stationary Gaussian process (\citealp{kammann2003geoadditive}), again originally specified for the anisotropy components, in which the vector of values $\boldsymbol{\phi} = \big(\phi(\bfs_1), \dots, \phi(\bfs_n) \big)$ is modeled as a linear combination of basis functions: 
\begin{equation} \label{approx_phi}
\log \boldsymbol{\phi} = \mu_\phi {\bf 1}_N + \sigma_\phi {\bf P}_\phi {\bf V}^{-1/2}_\phi {\bf w}_\phi.
\end{equation}
In (\ref{approx_phi}), ${\bf w}_\phi$ is a latent process defined on a set of knot locations $\{ {\bf b}_k: k = 1, \dots, K\}$, the radial basis functions ${\bf P}_\phi{\bf V}^{-1/2}_\phi$ are constructed using the $N \times K$ matrix of pairwise Mat\'ern correlations $C_\phi(\cdot; \rho_\phi, \nu_\phi)$ between the observation locations and knot locations (${\bf P}_\phi$) and the inverse square root of the $K\times K$ matrix of pairwise Mat\'ern correlations among the knot locations (${\bf V}_\phi$). If the knots correspond to the observation locations, this representation would correspond exactly with that in (\ref{sgp_phi}). The vector of unknown parameters includes ${\bf w}_\phi$ as well as $\{ \mu_\phi, \rho_\phi, \sigma_\phi, \nu_\phi \}$, but the dimension of the parameter space is now only $K+4$. Note that some of the hyperparameters $\{ \mu_\phi, \rho_\phi, \sigma_\phi, \nu_\phi \}$ might need to be fixed (see, e.g., \citealp{Paciorek2006}, Section 3.2.2). Again, note that \cite{Paciorek2006} use this approach for the scalar anisotropy components, but here we propose using the same statistical model for variance components.
 
Note that there is a correspondence between (\ref{approx_phi}) and the mixture component approach used in \cite{Risser2017}: certain values of the hyperparameters $\{ \mu_\phi, \rho_\phi, \sigma_\phi, \nu_\phi \}$ in (\ref{approx_phi}) make the two approaches (essentially) equivalent. For this paper we only include the approach given in \cite{Paciorek2006}.

\subsection{Matrix-valued anisotropy process} \label{Anis_proc}

The anisotropy process $\bfSig(\cdot)$ is defined over the space of $d \times d$ positive definite matrices. As such, it is less straightforward how to specify a process model for $\bfSig(\cdot)$. Several approaches are described in the following sections.

\subsubsection{Spatial constant}

Again, for completeness we define a spatially-constant anisotropy process:
\begin{equation} \label{constAniso}
\bfSig(\bfs) \equiv \bfSig, \hskip3ex \text{for all } \bfs \in G.
\end{equation}

\subsubsection{Covariance regression} \label{covreg}

Using the intuition from mean regression, \cite{Risser2015} use covariance regression (\citealp{Hoff2012}) to parameterize the anisotropy process. Specifically,
\begin{equation} \label{cr_formula}
\bfSig(\bfs) = {\bf \Psi} + {\bf \Gamma} \bfx_\Sigma(\bfs) \bfx_\Sigma(\bfs)^\top {\bf \Gamma}^\top,
\end{equation}
where $\bfx_\Sigma(\bfs) \in \mathcal{R}^{p_\Sigma}$ is a vector of relevant covariates (including an intercept, of length $p_\Sigma$), ${\bf \Psi}$ is a $d \times d$ positive definite matrix, ${\bf \Gamma}$ is a $d \times p_\Sigma$ real matrix. Because ${\bf \Psi}$ is positive definite, the number of parameters in this representation is $d(d+1)/2 + dp_\Sigma$.

\subsubsection{Componentwise regression} \label{compReg}

Alternatively, one might decompose the anisotropy process $\bfSig(\cdot)$ into the $d(d+1)/2$ unique processes. For example, \cite{Paciorek2006} suggest the eigendecomposition
\[
\bfSig(\bfs) = {\bf \Gamma}(\bfs) {\bf \Lambda}(\bfs) {\bf \Gamma}(\bfs)^\top,
\]
where ${\bf \Lambda}(\cdot)$ is a diagonal matrix of eigenvalues and ${\bf \Gamma}(\cdot)$ is a matrix of eigenvectors. Even in this representation, there are many ways to parameterize the unique parameters of each $\bfSig(\cdot)$ \citep[see, e.g.][]{Paciorek2006}; for $d=2$, we use
\begin{equation} \label{decomp}
{\bf \Lambda}(\bfs) = \left[ \begin{array}{cc} \lambda_1(\bfs) & 0 \\ 0 & \lambda_2(\bfs) \end{array} \right], \hskip3ex 
{\bf \Gamma}(\bfs) = \left[ \begin{array}{cc} \cos \gamma(\bfs) & -\sin \gamma(\bfs) \\ \sin \gamma(\bfs) & \cos \gamma(\bfs) \end{array} \right],
\end{equation}
as in \cite{Risser2015}, where we limit $\gamma(\bfs) \in \left[0, \frac{\pi}{2} \right]$ for identifiability. Now, in terms of $\lambda_1(\bfs)$, $\lambda_2(\bfs)$, and $\gamma(\bfs)$, we can define linear regression models on transformed versions of these parameters:
\begin{equation} \label{compReg_mod2}
\begin{array}{c}
\log\lambda_1(\bfs) = \bfx_\Sigma(\bfs)^\top \boldsymbol{\alpha}_{\lambda_1} \\[0.5ex]
\log\lambda_2(\bfs) = \bfx_\Sigma(\bfs)^\top \boldsymbol{\alpha}_{\lambda_2} \\[0.5ex]
\log \frac{\frac{2}{\pi}\gamma(\bfs)}{1 - \frac{2}{\pi}\gamma(\bfs)} = \bfx_\Sigma(\bfs)^\top \boldsymbol{\alpha}_{\gamma}, 
\end{array}
\end{equation}
where the transformations are such that each component has real support.

\subsubsection{Nonparametric regression}

Using the decomposition into three scalar processes given in (\ref{decomp}), we could also apply the approximate GP representation in (\ref{approx_phi}) to each of
\[
\left\{ \log\lambda_1(\cdot), \log\lambda_2(\cdot), \log \frac{\frac{2}{\pi}\gamma(\cdot)}{1 - \frac{2}{\pi}\gamma(\cdot)} \right\}. 
\]
Aside from their alternate parameterization, this is the approach used in \cite{Paciorek2006}.

\subsubsection{Local isotropy}

For spatial dimensions larger than $d=2$, estimating a spatially-varying anisotropy process becomes more challenging due to the large number of parameters needed to model the $d(d+1)/2$ unique processes in $\bfSig(\bfs)$. One (albeit simplified) way around this problem is to force the covariance to be locally isotropic by setting the anisotropy process to be equal to a multiple of the identity matrix:
\[
\bfSig(\bfs) \equiv \Sigma(\bfs){\bf I}_d,
\]
where $\Sigma(\bfs)$ is a scalar. In this case, Equations (\ref{PScov}) and (\ref{Qij}) become
\[
Q(\bfs, \bfs') = \frac{(\bfs - \bfs')^\top (\bfs - \bfs')}{\frac{1}{2}[\Sigma(\bfs) + \Sigma(\bfs')] } = \frac{||\bfs - \bfs'||^2}{\frac{1}{2}[\Sigma(\bfs) + \Sigma(\bfs')]},
\]
and
\begin{equation} \label{iso_cov}
C_y(\bfs, \bfs'; \bftheta) = \sigma(\bfs) \sigma(\bfs') \frac{\big(\Sigma(\bfs)\Sigma(\bfs')\big)^{d/4}}{\left(\frac{\Sigma(\bfs') + \Sigma(\bfs')}{2} \right)^{d/2}} g\left(\sqrt{Q(\bfs, \bfs')}\right), \hskip3ex \bfs, \bfs' \in G.
\end{equation}
Note that now $Q(\bfs, \bfs')$ (and hence $C_y$) only depends on the squared Euclidean distances $||\bfs - \bfs'||^2$. In this case, we can apply any of the componentwise and nonparametric regression frameworks to $\log \Sigma(\bfs)$ to estimate the nonstationary but locally isotropic (instead of locally stationary) covariance function in (\ref{iso_cov}) for an arbitrary spatial dimension $d \geq 1$.

\section{Approximate inference and prediction for large data} \label{section4}

Despite the fact that Gaussian processes are mathematically convenient representations for a spatial process and that prediction is straightforward, numerical calculations regarding the multivariate Gaussian distribution for $N$ spatial locations require $\mathcal{O}(N^2)$ memory and $\mathcal{O}(N^3)$ time complexity. This is an issue for any application of Gaussian processes, but is particularly problematic for modeling nonstationary covariance functions which involve high-dimensional parameter spaces. However, when dealing with large data sets, we can utilize the diverse literature on approximate Gaussian process methods \citep[see][for a review and comparison of existing approaches]{heaton2018case} to make parameter inference and prediction feasible. 

The nearest-neighbor Gaussian process (NNGP; \citealp{datta2016hierarchical}) and the sparse general Vecchia (SGV; \citealp{katzfuss2017general} and \citealp{katzfuss2018vecchia}) approximations are two specific methods that enable large data inference via Gaussian processes by forcing the precision matrix to be sparse. 
Both of these methods can be framed as special cases of Vecchia approximations of Gaussian processes (\citealp{katzfuss2017general}); for now focusing on the distribution of the latent process $y(\cdot)$ at the observed locations $\mathcal{S}_O$, note that we can write (\ref{Ycond}) as 
\begin{equation} \label{GPcond}
p({\bf y}_O) = p(y_1) \prod_{i=2}^N p\big(y_i | {\bf y}_{h(i)} \big),
\end{equation}
(implicit conditioning on $\bftheta$ and $\bfbeta$ suppressed for simplicity) where we denote $y_i = y(\bfs_i)$, $h(i) = (1, \dots, i-1)$, and ${\bf y}_{h(i)} = \{y_j: j \in h(i)\}$; note that in this framework the ordering of the locations in $\mathcal{S}_O$ is assumed to be arbitrarily fixed. However, (\ref{GPcond}) provides no computational shortcuts, as the conditional densities still involve $\mathcal{O}(N^2)$ memory and $\mathcal{O}(N^3)$ computations. Vecchia's approximation (\citealp{vecchia1988estimation}) suggests an approximation to (\ref{GPcond}) wherein the conditioning sets $h(i)$ are replaced with subvectors $g(i) \subset h(i)$. We utilize terminology from \cite{katzfuss2017general} and refer to $g(i)$ as the $i$th conditioning index vector and ${\bf y}_{g(i)}$ as the conditioning vector for $y_i$, yielding the Vecchia approximation of the joint density in (\ref{GPcond}):
\begin{equation}\label{GPapprox}
\hat{p}({\bf y}_O) = p(y_1) \prod_{i=2}^N p\big(y_i | {\bf y}_{g(i)} \big).
\end{equation}
The Vecchia approximation (\ref{GPapprox}) converges to the true distribution in (\ref{GPcond}) as the conditioning vectors $g(i)$ approach $h(i)$, but of course we prefer small conditioning vectors for computational efficiency.

A more general framework applies to the vector ${\bf w}_O = {\bf y}_O \cup {\bf z}_O$, where (again following notation from \citealp{katzfuss2017general}) the ordering in ${\bf w}_O$ is such that the $y_i$ retain their relative ordering in ${\bf w}_O$ and the $z_i$ are inserted directly after $y_i$. The general Vecchia approximation (\citealp{katzfuss2017general}) can now be written as 
\begin{equation} \label{genVecc2}
\hat{p}({\bf w}_O) = \prod_{i=1}^N \Big[ p\big(y_i | {\bf y}_{q_y(i)}, {\bf z}_{q_z(i)} \big) \times p(z_i | y_i) \Big],
\end{equation}
(where we now define $g(1) = \emptyset$). The conditioning vector for $z_i$ is always $y_i$ because (\ref{CANONmodel}) assumes $z_i$ is conditionally independent of all other observed responses given $y_i$. In (\ref{genVecc2}), the conditioning vector for $y_i$ has been split into two sub-vectors $q_y(i)$ and $q_z(i)$, where $j\in q_y(i)$ means $y_i$ conditions on $y_j$ and $j\in q_z(i)$ means $y_i$ conditions on $z_j$ (in other words, does $y_i$ condition on the latent or observed value). Finally, we assume $q_y(i) \cap q_z(i) = \emptyset$ and denote $q(i) = (q_y(i), q_z(i))$.

\cite{katzfuss2017general} state that a general Vecchia approximation $\hat{p}({\bf w}_O)$ of $p({\bf w}_O)$ is determined the ordering of the locations in $\mathcal{S}_O$, the conditioning index vector $q(i) \subset (1, \dots, i-1)$ for $y_i$, and partitioning $q(i)$ into $q_y(i)$ and $q_z(i)$. (Note that \citealp{katzfuss2017general} further describe the choice of a superset of the observations and a partitioning the superset; we simply use default choices of setting the superset equal to $\mathcal{S}_O$ and using a scalar partitioning.) Regarding the first choice, we use an approximate maximum-minimum-distance (maxmin) ordering (\citealp{guinness2018permutation}), which has been shown as optimal for $d\geq 2$ dimensional spatial domains (\citealp{katzfuss2017general}). Regarding the second choice, we use $k$ nearest-neighbor (NN) conditioning, i.e., $q(i) = (1, \dots, i-1)$ for $i \leq k$ and $q(i)$  consists of the $k$ locations from $(1, \dots, i-1)$ with the smallest Euclidean distance from $\bfs_i$ otherwise. The third choice has a greater impact on the computational properties of the resulting Vecchia approximation (\ref{genVecc2}), and we present two cases in Sections \ref{sec41} (sparse general Vecchia or SGV) and \ref{sec42} (nearest neighbor GP for the response or NNGP-R).

\subsection{Sparse general Vecchia} \label{sec41}

The sparse general Vecchia (SGV) approximation partitions the conditioning index vector as described in Section 5 of \cite{katzfuss2017general}. The SGV partitioning ensures that the corresponding directed acyclic graph forms a ``perfect graph'' \citep{lauritzen1996graphical} and provides a compromise between conditioning only on the $y(\cdot)$ (which provides a better approximation but is slower computationally) and only on $z(\cdot)$ (which provides a worse approximation but is much faster; see Section \ref{sec42}). The general Vecchia approximation implied by (\ref{genVecc2}) is the multivariate Gaussian distribution $N_n(\boldsymbol{\mu}_O, {\bf \Sigma}_O)$. The SGV guarantees that both ${\bf \Sigma}_O^{-1}$ and ${\bf U}_O$ are sparse, where ${\bf \Sigma}_O^{-1} = {\bf U}_O {\bf U}_O^\top$ and ${\bf U}_O$ is the upper triangular Cholesky factor based on a reverse row-column ordering of ${\bf \Sigma}_O^{-1}$ (\citealp{katzfuss2017general}). For a known covariance function and a fixed value of the covariance parameters, a closed form expression is available for the non-zero elements of ${\bf U}_O$ \citep[Proposition 1,][]{katzfuss2017general}; thus, ${\bf U}_O$ has at most $k+1$ nonzero entries per column, such that ${\bf U}_O$ can be computed in $\mathcal{O}(Nk^3)$ time \citep{katzfuss2017general}. In order to use this likelihood approximation in a Markov chain Monte Carlo (MCMC) algorithm, we require the approximate marginalized likelihood of ${\bf z}_O$; a closed form expression in terms of the mean and covariance parameters is given in Proposition 2 of \cite{katzfuss2017general}. The time complexity for computing the approximate marginalized likelihood is $\mathcal{O}(Nk^2)$ \citep[][Proposition 6]{katzfuss2017general}; thus, SGV approximation retains linear computational complexity in $N$. 

Prediction in the SGV approach can be framed as a post-processing step when adopting an ordering scheme that first orders the observed locations and then orders the prediction locations \citep[the so-called ``obs-pred'' ordering; see][]{katzfuss2018vecchia}. While this restricts the choice of ordering somewhat, the primary benefit is that under obs-pred ordering the approximate likelihood for just ${\bf w}_O$ is equal to the marginalized likelihood of the combined vector $({\bf w}_O, {\bf y}_P)$ (integrated with respect to ${\bf y}_P$); thus, likelihood inference can first be carried out based on just ${\bf w}_O$ and then ${\bf y}_P$ can be appended when predictions are desired, without changing the distribution $\hat{p}({\bf w}_O)$ \citep{katzfuss2018vecchia}. As with likelihood inference, predictions for SGV under obs-pred ordering can be obtained in $\mathcal{O}(Nk)$ time complexity, which is linear in $N$; see Section 3.4 of \cite{katzfuss2018vecchia} for more information on conditional simulation.

\subsection{Nearest Neighbor Gaussian Process-Response} \label{sec42}

The nearest neighbor Gaussian process for the response \citep{finley2018efficient} can be seen as a special case of the general Vecchia approximation where the conditioning index vector for all $y_i$ includes only elements of the observed ${\bf z}_O$; i.e., $q_y(i) = \emptyset$. Given that the NNGP-R approximation is a special case of the general Vecchia approximation, likelihood calculations could be conducted as with SGV. However, given the conditioning structure, with NNGP-R one can explicitly integrate over the ${\bf y}_O$ and apply the likelihood approximation directly to $p({\bf z}_O)$ (from Eq.~\ref{Zmarg}; conditioning on mean and covariance parameters suppressed), whereas the general Vecchia framework specifies an approximation to $p({\bf w}_O) = p({\bf y}_O, {\bf z}_O)$. The NNGP-R approximation is now applied to yield a sparse Cholesky factor of $\text{Cov}({\bf z}_O)$, which is based on $C_z$ (from Eq.~\ref{Zcov}), as opposed to the general Vecchia which calculates the sparse Cholesky factor of $\text{Cov}({\bf y}_O, {\bf z}_O)$ based on $C_y$ and $\tau(\cdot)$. Furthermore, \cite{finley2018efficient} derive a closed form expression for calculating the Cholesky factor of $\text{Cov}({\bf z}_O)$ and the subsequent quadratic forms needed to evaluate the likelihood. When the number of nonzero elements in the Cholesky decomposition is limited to $k$ (again using a $k$ nearest neighbor scheme based on maxmin ordering), the Cholesky is guaranteed to be sparse and can be calculated by solving $N-1$ linear systems of size at most $k\times k$, which can be performed serially in $\mathcal{O}(Nk^3)$ flops. As such, the likelihood can then be calculated in $\mathcal{O}(Nk)$ time complexity \citep{finley2018efficient}, which is linear in $N$.

The tradeoff of conditioning on the observed ${\bf z}_O$ is that posterior prediction can only be accomplished for individual locations (also called ``local kriging'') because the covariance corresponding to the prediction locations is diagonal \citep[Section 5.2.1 of][]{katzfuss2018vecchia}. \cite{finley2018efficient} outline an algorithm for posterior prediction of the response (i.e., $z(\cdot)$) at a single location \citep[Algorithm 4,][]{finley2018efficient}; \cite{katzfuss2018vecchia} note that the same framework can be used to predict either $z(\cdot)$ or $y(\cdot)$ by including or not including the nugget variance, respectively, in the prediction variance. 

\subsection{Comparing SGV and NNGP-R}

In summary, while the SGV and NNGP-R approximations both arise as special cases of the general Vecchia approximation, there are important differences in the underlying properties of the resulting approximation. First, the approximation accuracy is better for SGV, relative to NNGP-R \citep[Proposition 4 in][]{katzfuss2018vecchia}. However, \cite{katzfuss2018vecchia} show empirically that both likelihood calculations and prediction are much faster for NNGP-R relative to SGV \citep[Figure 5,][]{katzfuss2018vecchia}. SGV performs better in the low signal-to-noise situation; also, SGV can characterize joint predictions whereas NNGP-R can only yield marginal (univariate) predictions. This last feature is potentially the most problematic, since joint predictions are required for uncertainty quantification of spatial averages as well as generating statistical ensembles of the underlying process of interest \citep[see, e.g.,][]{li2019efficient}. This important trade-off between computational speed and approximation accuracy/joint prediction is the primary reason why we have chosen to include both of these methods in this paper (and the corresponding software package), as the specific application may motivate a preference for speed vs. accuracy, and furthermore if joint predictions are required or not.

\section{Implementation} \label{implem}

The methods outlined in Sections \ref{CGPM}, \ref{section3}, and \ref{section4} are implemented in the \pkg{BayesNSGP} package (version 0.1.1), now available on CRAN (Comprehensive R Archive Network), the central repository for curating and disseminating R packages \citep{hornik2012comprehensive}. Here, we briefly describe the two outward-facing user interface functions, \texttt{nsgpModel} and \texttt{nsgpPredict}, which respectively fit a Bayesian spatial Gaussian process and generate posterior predictions for a set of locations. An expanded explanation of the package is available as a vignette in the online Supplemental Materials.

First, the \texttt{nsgpModel} function enables MCMC for a general nonstationary spatial GP. The structure of a particular spatial model is defined by choices of \texttt{tau\char`_model}, \texttt{sigma\char`_model}, \texttt{Sigma\char`_model}, and \texttt{mu\char`_model}, which respectively specify statistical models for $\tau(\cdot)$, $\sigma(\cdot)$, $\Sigma(\cdot)$, and the mean of the process $\mu(\cdot)$. Any or all of the parameter processes can utilize the various statistical models outlined in Section \ref{section3}; the syntax for each option is provided in the online Supplemental Materials (e.g., \texttt{tau\char`_model = "logLinReg"} uses log-linear regression for the $\tau(\cdot)$ process).
The \texttt{likelihood} argument specifies the likelihood to use--choosing from \texttt{"fullGP"} (the exact Gaussian likelihood), \texttt{"SGV"} (the sparse general Vecchia), or \texttt{"NNGP"} (the nearest-neighbor Gaussian process). The user is also required to specify a $N\times d$ matrix of spatial coordinates, the $N$-vector of measurements, and a list of constants (e.g., design matrices, fixed data-level hyperparameters, and fixed prior-level parameters) needed to build and compile the \texttt{nimbleModel} object, which will later be used to compile and run the Markov chain Monte Carlo (again see the online Supplemental Materials). The object returned from \texttt{nsgpModel} is indeed a \pkg{nimble} ``model'' object, which is \pkg{nimble}'s abstraction for a hierarchical statistical model.  \pkg{nimble} models provide the ability to store values into model parameters or latent states, designate values as observed data, simulate new values from prior distributions, and calculate log-densities.  In addition, \pkg{nimble} algorithms (for example, \pkg{nimble}'s MCMC) natively operate on \pkg{nimble} model objects.

The general workflow is first creating a \pkg{nimble} model, then creating an MCMC algorithm to fit this model.  The MCMC can optionally be customized, for example by assigning different samplers, such as slice sampling \citep{neal2003slice}, Metropolis-Hastings sampling on a log scale for variance components, or joint sampling of correlated parameters using multivariate Metropolis-Hastings.  Details of the sampling algorithms available with \pkg{nimble} can be found in \proglang{R} using \texttt{help(samplers)}, or in the \pkg{nimble} User Manual (\texttt{https://r-nimble.org/manuals/NimbleUserManual.pdf}).  Once ready, both the model and MCMC are compiled to C++ (functionality provided by \pkg{nimble}) for faster execution.  Finally, the compiled MCMC algorithm is executed to generate posterior samples.

Next, the \texttt{nsgpPredict} function enables posterior prediction as a straightforward post-processing step for any of the likelihood methods via posterior samples generated using a \texttt{nsgpModel} object and the \pkg{nimble} package. The user must provide a matrix of prediction coordinates and corresponding constants for the prediction locations, as well as indicate whether prediction should correspond to the $y(\cdot)$ process or $z(\cdot)$. When the necessary constants are provided, \texttt{nsgpPredict} proceeds to conduct posterior prediction as described in Sections \ref{CGPM} and \ref{section4}, depending on the likelihood model. 

Three additional helper functions are necessary for setting up the required constants for the \texttt{nimbleModel} when using the SGV and NNGP likelihood models: \texttt{orderCoordinatesMMD}, \texttt{determineNeighbors}, and \texttt{sgvSetup}. All of these functions operate internally within a call of the \texttt{nsgpModel} function, but it is important to understand what these functions are doing in the background. For both NNGP and SGV, the first step is to re-order the coordinates (and all corresponding quantities, e.g., the $z(\cdot)$ values, design matrices, etc.) following an approximate maximum-minimum distance \citep[MMD;][]{guinness2018permutation} ordering; this is accomplished using \texttt{orderCoordinatesMMD}. Next, the approximate likelihood methods require a set of neighbors for each location; \texttt{determineNeighbors} takes a set of spatial coordinates (of size $N\times d$) as well as the desired number of neighbors, $k$, and returns a $N \times k$ matrix indexing the coordinates that are nearest neighbors for each location. Recall that the order of the coordinates matters -- so, for coordinate $i = 1, \dots, N$, \texttt{determineNeighbors} finds the $k$ nearest coordinates (in terms of Euclidean distance) from the preceding $i-1$ coordinates. A necessary implication of this is that the first coordinate has no neighbors and coordinates $i = 2, \dots, k$ only have $i-1$ neighbors. The wrapper \texttt{sgvSetup} function conducts three operations that are required for the SGV likelihood: the re-ordering and neighbor identification already described, as well as a determination of the conditioning sets. Recall from Section \ref{section4} that for location $i = 1, \dots, N$, we require the conditioning set $q(i)$ -- i.e., for each neighbor $j$, whether we should condition on $y(\bfs_j)$ or $z(\bfs_j)$. 

\section{Applications} \label{section6}

In order to demonstrate our methods, we now analyze two data sets. The code used to conduct each of the following analyses is available in the Supplemental Materials.

\subsection{Annual precipitation for Colorado in 1981} \label{application1}

\begin{figure}[!t]
\begin{center}
\includegraphics[trim={0 0 0 0mm}, clip, width = \textwidth]{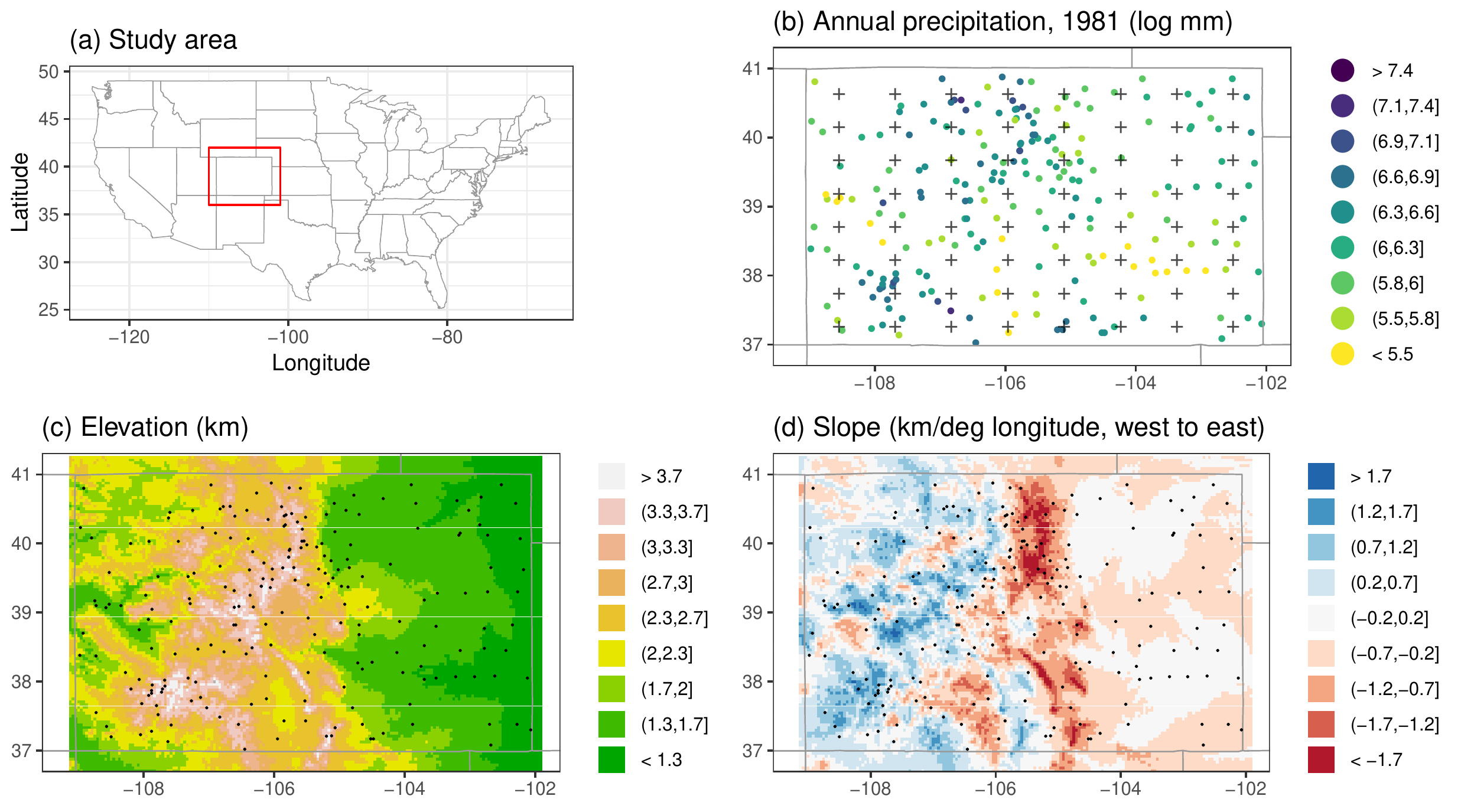}
\caption{The area of study (panel a), with annual precipitation totals for weather stations in 1981 (panel b, in log mm). Elevation (km) and west-to-east gradient (km per $^\circ$longitude) shown in panels (c) and (d), with the weather station locations overlaid. Panel (b) also includes the knot locations of the latent processes for $\bfSig(\cdot)$.}
\label{figure1}
\end{center}
\end{figure}

First, we reproduce an analysis of the 1981 annual precipitation data set used by \cite{Paciorek2006} with $N = 217$ measurements from Colorado, a state in the western United States of America. Annual precipitation totals are given in millimeters, and we analyze the log of total precipitation to make the Gaussian process assumption more appropriate (see Figure \ref{figure1}). Figure \ref{figure1} also shows the diverse topography in Colorado (panel c) as well as a derived measure of the change in elevation (``slope''; panel d), measured as a west-to-east gradient. In panel (d), dark blue colors indicate steep west-facing mountainsides; white indicates relatively flat terrain; and dark red colors indicate steep east-facing mountainsides. The specific analysis is an illustration of two previous analyses of this data set: (1) the original analysis in \cite{Paciorek2006}, and (2) a subsequent analysis using the regression-based nonstationary covariance function from \cite{Risser2015}. Both of these analyses are special cases of our general methodology, corresponding to specific choices and combinations of the various submodels in Section~\ref{section3}. Our goal is to demonstrate how these models, both of which use the exact GP likelihood, can be quickly implemented using the \pkg{BayesNSGP} package and \pkg{nimble}.

\subsubsection{Original analysis} \label{app1Orig}

In \cite{Paciorek2006}, the mean $\mu(\bfs)$, the nugget variance $\tau(\bfs)$, and the spatial variance $\sigma(\bfs)$ were modeled as unknown constants, while the anisotropy process $\Sigma(\bfs)$ was modeled using an approximation to a Gaussian process. The following aspects of the analysis are designed to match \cite{Paciorek2006}: first, we use a coarse, evenly spaced $8 \times 8$ grid of $K=64$ knot locations; next, we fix the smoothness of the latent GPs to be $\nu_\phi = 5$. While \cite{Paciorek2006} assign a uniform prior on the log scale to the latent GP range, with upper bound $\log 3.85$ and lower bound $\log 0.1$, we instead use a $U(0, 3.85)$ prior on this parameter. In the original paper they estimate the data-level smoothness with a $U(0.5, 30)$ prior while we fix this smoothness parameter at $\nu = 2$; finally, we set an upper bound on the eigenvalue processes to be 16. Note that unlike \cite{Paciorek2006} we estimate the mean and standard deviation of the latent GPs (i.e., $\mu_\phi$ and $\sigma_\phi$): the GP means are assigned diffuse, mean-zero Gaussian priors; the GP standard deviations are assigned uniform priors over the interval from zero to 10 and 20 (respectively, for the two eigenvalue processes and the rotation process). Finally, the spatial mean is assigned a diffuse, mean-zero Gaussian prior.
The appropriate \texttt{nimble} model can then be created and configured using \texttt{tau\char`_model = "constant"}, \texttt{sigma\char`_model = "constant"}, \texttt{Sigma\char`_model = "npApproxGP"}, and \texttt{mu\char`_model = "constant"}. 

The primary complication with the \cite{Paciorek2006} methodology is that the posterior is very difficult to sample from using MCMC: the model as described here requires sampling from a 201-dimensional posterior (with only 217 measurements), and the $\{w_1(\cdot), w_2(\cdot), w_3(\cdot)\}$ latent anisotropy processes are likely both auto-correlated and cross-correlated. The \pkg{nimble} functionality makes it very easy to test various schemes for block sampling, particularly for the latent anisotropy processes, where it is not clear how to best sample the $64\times 3 = 192$ parameters. We tested the 16 sampling schemes detailed in Table \ref{app1_time}, which explore different sampling schemes for the latent $\{w_1(\cdot), w_2(\cdot), w_3(\cdot)\}$ and their hyperparameters $\{ \mu_\Sigma^k, \sigma_\Sigma^k, \phi_\Sigma^k: k = 1, 2\}$. These schemes consist of combinations of the following: sampling spatial sub-blocks of the $\{w_j(\cdot): j = 1,2,3\}$ of size 4, 8, 16, or 32; sampling the $\{w_j(\cdot): j = 1,2,3\}$ jointly (meaning we actually have blocks of 12, 24, 48, or 96) or separately; and using random walk Metropolis-Hastings or slice samplers for the latent hyperparameters. Otherwise, for all of these schemes, we use univariate Metropolis-Hastings samplers for the spatially-constant mean, nugget variance, and spatial variance.  Each MCMC was run for 100,000 iterations, and we discard the first 50,000 as burn-in and save every 10$^\text{th}$ sample (for a total of 5,000 posterior samples). The various MCMC sampling schemes (see Table \ref{app1_time} for full details) are evaluated in terms of their minimum efficiency, which is defined as the minimum effective sample size across all 201 parameters divided by the run time (see Section \ref{app1Results} for results). Finally, we conduct posterior prediction for a fine grid over Colorado using the 5,000 thinned post burn-in posterior samples.

\subsubsection{Subsequent analysis} \label{app1_subseq}

Using the regression-based analysis of \cite{Risser2015}, covariate information can be included in three parts of the nonstationary model: the mean function, the spatial variance function, and the kernel matrix function. \cite{Risser2015} consider elevation as well as a covariate that describes change in elevation (``slope''; see Figure \ref{figure1}). Here, we reproduce their FNS-M2 implementation, which includes the main effects of elevation and slope as well as their interaction in each of the mean, variance, and kernel matrix functions. As mentioned above, following \cite{Risser2015}, elevation and slope measurements are standardized in order to put the coefficient estimates on a similar scale. Setting up the model as before, we use the exponential correlation function \citep[as in][]{Risser2015} and the same prior hyperparameters. The appropriate \texttt{nimble} model can then be created and configured using \texttt{tau\char`_model = "constant"}, \texttt{sigma\char`_model = "logLinReg"}, \texttt{Sigma\char`_model = "covReg"}, and \texttt{mu\char`_model = "linReg"}. 
As in \cite{Risser2015}, we implement adaptive univariate random walk samplers for all parameters except for a block Metropolis Hastings sampler for the parameters in ${\bf \Psi}$, and run a total of 100,000 iterations of the MCMC (discarding the first 50,000 as burn-in) followed by posterior prediction using a thinned chain (every 10$^\text{th}$ sample) of post burn-in posterior samples. 

\subsubsection{Results} \label{app1Results}

\begin{figure}[!t]
\centering
\includegraphics[width=\textwidth]{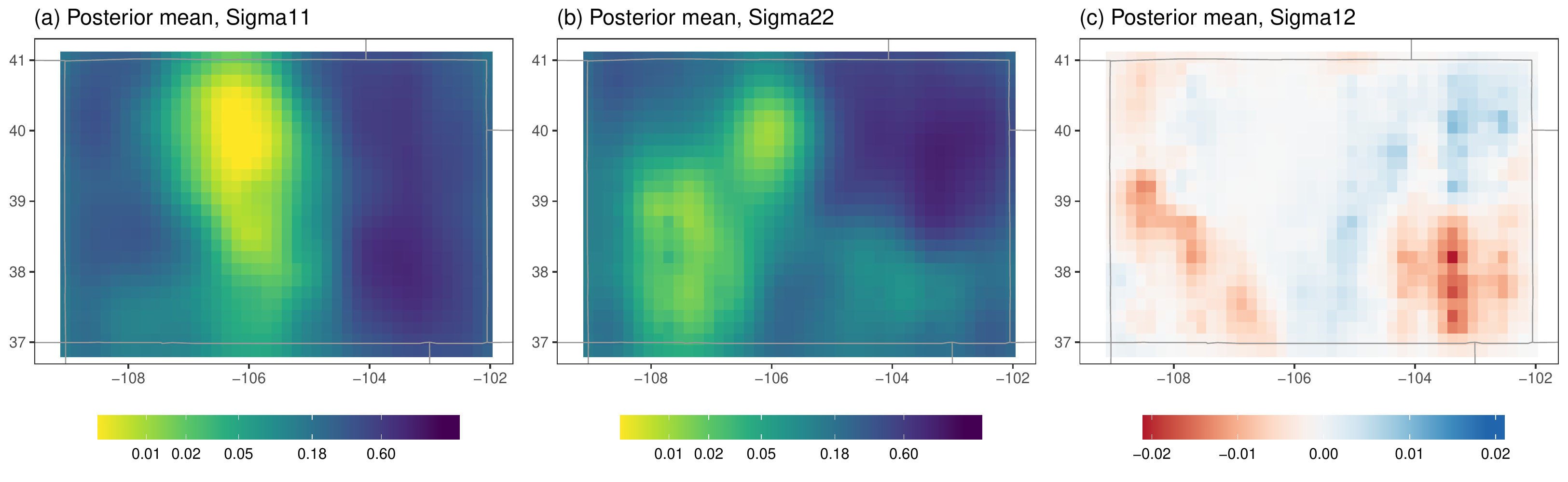} 
\includegraphics[trim={0 45 0 5mm}, clip, width = 0.6\textwidth]{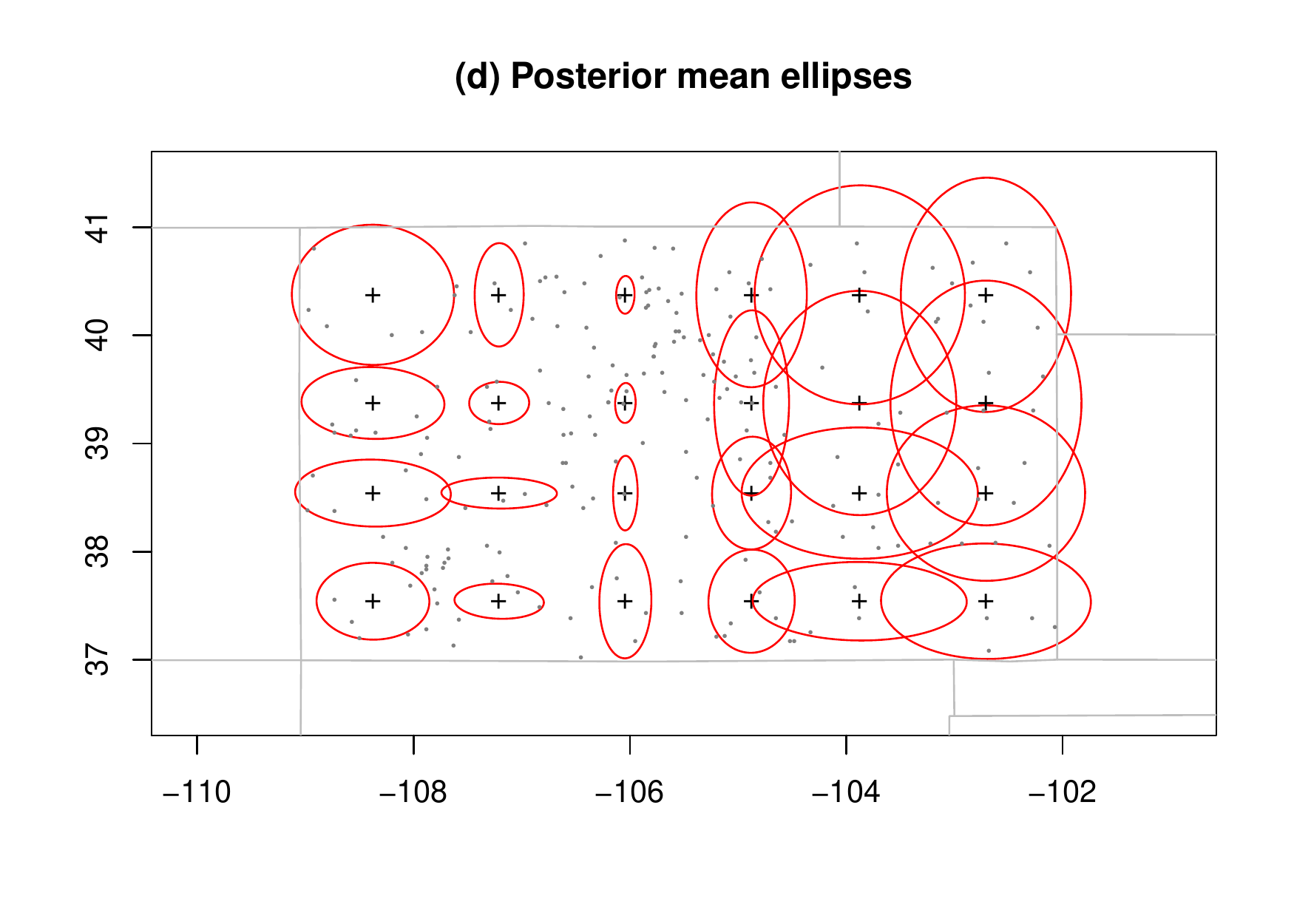} 
\caption{Panels (a), (b), and (c) contain spatial maps of the posterior mean of the anisotropy process $\bfSig(\cdot)$ elements (namely $\Sigma_{11}(\cdot)$, $\Sigma_{22}(\cdot)$, and $\Sigma_{12}(\cdot)$, respectively). To simultaneously view the effect of these parameter processes, panel (d) shows the posterior mean 50\% probability ellipse of a bivariate Gaussian density with covariance matrix $\bfSig(\cdot)$ for a set of representative locations over the domain \citep[compare with Figure 4d in][]{Paciorek2006}.}
\label{Sigma_plots}
\end{figure}

In spite of the MCMC for the \cite{Paciorek2006} analysis being notoriously difficult to run, the sub-block random walk samplers are effective at sampling the highly correlated, high-dimensional posteriors for each of $\{w_1(\cdot), w_2(\cdot), w_3(\cdot) \}$. Table \ref{app1_time} shows the minimum efficiency for all 16 of the sampling schemes considered, which appears to favor Scheme 7 or 8. However, after examining the posterior distributions for the spatial fields of the latent anisotropy processes, Schemes 7, 8, 15, and 16 were excluded from consideration based on poor mixing of the MCMC (which is not entirely surprising, since these schemes involve block sampling of 48 or 96). Otherwise, Scheme 6 is identified as best in terms of minimum efficiency, and we use the posterior draws generated using this sampling scheme for all subsequent analysis. The MCMC for Scheme 6 is relatively fast ($\approx 13$ hours for 100,000 iterations); returning to the subsequent analysis in Section \ref{app1_subseq}, the MCMC for the \cite{Risser2015} analysis took just 4.4 hours for 100,000 iterations, which is significantly faster than (at least) the analysis in \cite{Risser2015}, which the authors mention took approximately 20 hours for 10,000 iterations (although the compute machines are not directly comparable).

The estimated anisotropy process $\bfSig(\cdot)$ for the \cite{Paciorek2006} analysis is somewhat difficult to summarize given the high dimensionality of the parameters involved; however, spatial maps of the posterior mean of the elements of $\bfSig(\cdot)$ are shown in Figure \ref{Sigma_plots}(a)-(c), with a qualitative summary of the spatially-varying magnitude and direction of spatial dependence in Figure \ref{Sigma_plots}(d). The ellipses in panel (d) represent the posterior mean 50\% probability ellipse of a bivariate Gaussian density with covariance matrix $\bfSig(\cdot)$ for a set of representative locations over the domain. This figure is comparable to Figure 4d \cite{Paciorek2006}: note that as in the original analysis, the ellipses (and hence range of dependence) are much smaller in the center of the domain--where there is highly diverse topography--but large in the eastern part of Colorado where the topography is flat, and also somewhat larger in the far western part of the state where the topography is much less heterogeneous.

Next, we can compare posterior summaries for all statistical parameters in the covariance regression analysis of \cite{Risser2015}; specifically, comparing the posterior means and 95\% Bayesian credible intervals taken from Table 1 of \cite{Risser2015} with the results obtained from our analysis using \pkg{BayesNSGP}. This comparison is provided in Figure \ref{compareEnv}: clearly, the \pkg{BayesNSGP} analysis reproduces the results from the original analysis.

Finally, we provide posterior prediction maps (both the mean and standard deviation) for both analyses in Figure \ref{app1_postPred} (these plots are not provided for \citealp{Risser2015}; the two left panels can be compared with Figure 3 in \citealp{Paciorek2006}). The two analyses using \pkg{BayesNSGP} yield similar posterior mean prediction maps, although the map for \cite{Risser2015} is much less smooth than \cite{Paciorek2006} (likely a function of both using the exponential instead of the Mat\'ern as well as including topographic covariates in the mean). Furthermore, the standard errors for the \cite{Paciorek2006} model are almost uniformly larger, particularly in the western part of the domain.

In summary, in this section we have demonstrated how our methods and corresponding software package can quickly and efficiently reproduce the results of two separate analyses of the same data set using a nonstationary covariance function with an exact Gaussian process likelihood. Furthermore, the customization of \pkg{nimble}'s MCMC allows us to quickly implement a variety of MCMC sampling schemes to arrive at an effective configuration for sampling from a notoriously difficult posterior distribution.

\subsection{Total precipitation over CONUS} \label{application2}

\begin{figure}[!t]
\begin{center}
\includegraphics[trim={0 0 0 0mm}, clip, width = \textwidth]{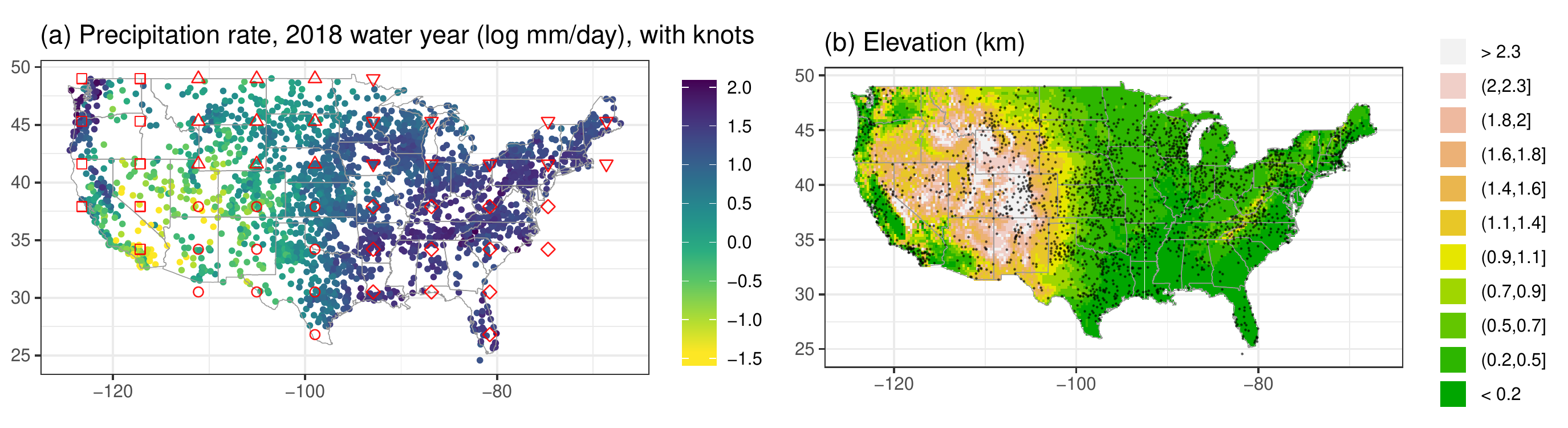}
\caption{Log precipitation rate during the 2018 water year (October 1, 2017 to September 30, 2018) for the $N=2311$ GHCN stations with no missing daily measurements over this time period (log mm day$^{-1}$; panel a), and elevation (km; panel b) with the station locations overlaid. Panel (a) also shows the knot locations, where the plotted shape represents the sampling sub-block.}
\label{app2_data}
\end{center}
\end{figure}

Next, we analyze a larger data set consisting of measurements of the log daily precipitation rate from the Global Historical Climatology Network-Daily database \citep[GHCN-D;][]{ghcnd_data, Menne2012} over the contiguous United States (CONUS) for the 2018 water year (October 1, 2017 to September 30, 2018). A total of $N=2311$ GHCN-D stations (out of 21,269 total; see Figure~\ref{app2_data}) have no missing daily values over this time period. Specifically, we are interested in analyzing the average daily precipitation over this water year (shown in Figure \ref{app2_data}), modeled on the log scale. The CONUS, particularly its western half, is a highly heterogeneous spatial domain with variable topography interacting with a diverse set of physical phenomena that produce precipitation, including atmospheric rivers, extratropical cyclones, tropical cyclones, and mesoscale convective systems. As such, it is important to fit a nonstationary spatial model to these data.

\begin{table}[!t]
\caption{A summary of the models fit to the log daily precipitation rate data (GP = Gaussian process). All models use a constant mean and an underlying exponential correlation function.}
\begin{center}
\begin{tabular}{|p{2cm}|p{7.7cm}|p{2.35cm} |}
\hline
\textbf{Label} 	& \textbf{Details} 			& \textbf{\proglang{R} package} \\ \hline\hline 
TGP$^\dagger$	& Bayesian treed GP			& \pkg{tgp} \\ \hline 
IGP$^\dagger$	& Bayesian isotropic GP 	& \pkg{spBayes}  \\ \hline 
PP$^\dagger$	& Bayesian predictive process ($r = 237$ knots) & \pkg{spBayes}	\\ \hline 
LSGN$^{\dagger *}$	& Locally stationary, globally nonstationary GP (with 41 mixture component locations)&  \pkg{convoSPAT}	\\ \hline \hline
AGP-SGV	        & Bayesian anisotropic GP with SGV likelihood approximation	&  \pkg{BayesNSGP} \\ \hline
NGP-SGV	        & Bayesian nonstationary GP with SGV likelihood approximation	&  \pkg{BayesNSGP} \\ \hline
\end{tabular}
\end{center}
\label{modelComp}
\begin{flushleft}
\vskip-2ex
{\scriptsize $^\dagger$Methods using pre-existing off-the-shelf software.}\\
{\scriptsize $^*$Implemented in a Frequentist framework.}\\
\end{flushleft}
\end{table}%

As described in Section \ref{section1}, there are a small number of \proglang{R} packages that provide fully Bayesian analysis of spatial data with a nonstationary covariance function. With this data set, we can compare these existing methods with our novel methods and the \pkg{BayesNSGP} package. Table \ref{modelComp} summarizes these methods and their software, as well as several statistical models fit using our new methodology (all described below). For completeness, we also include isotropic and stationary models in our comparison. In order to focus on the second-order properties of the fitted model, all models use a spatially-constant mean. The MCMC for all models is run for 40,000 iterations, discarding the first 30,000 as burn-in, and prediction is conducted for every tenth post burn-in sample (yielding 1,000 posterior predictive draws for each model). All models use an underlying exponential correlation function, and all use uninformative (most often uniform) priors on model parameters. For each model, we assess three comparisons: (1) the computational time required to run the MCMC and its corresponding efficiency, (2) a quantitative evaluation of model quality using 10-fold cross-validation, and (3) a qualitative assessment of the resulting posterior predictive means and standard deviations.

\vskip2ex
\noindent {\textit{Treed Gaussian process}}. {The treed Gaussian process (TGP) model of \cite{Gramacy2008} obtains a nonstationary covariance functions by partitioning the domain into multiple segments such that the process is independent across segments and stationary within the segments. Uncertainty in the tree partitioning is account for via Bayesian model averaging \citep{BMA1999}. The TGP model is implemented in the \pkg{tgp} package for \proglang{R} \citep{Gramacy2007}; we use all default settings, except for using a constant mean function.}

\vskip2ex
\noindent {\textit{Bayesian isotropic Gaussian process}}. 
{For comparison, we fit a Bayesian isotropic Gaussian process to these data using the \pkg{spBayes} package for \proglang{R} \citep{Finley2007}.}

\vskip2ex
\noindent {\textit{Gaussian predictive process}}. {Given the moderately large size of the precipitation rate data, we consider an off-the-shelf implementation of the Gaussian predictive process (PP) of \cite{Banerjee2008}, which is designed to handle large data sets. The PP constructs an approximation to a Gaussian process by projecting realizations of the process of interest onto a lower dimensional space spanned by a set of knot locations, which reduces the computational requirements of likelihood evaluations. The PP model is technically nonstationary by construction, but we note that its implementation in \pkg{spBayes} \citep{Finley2007}  has only trivial differences from a stationary covariance function. We fit the PP model using \pkg{spBayes} with 237 knots.} 

\vskip2ex
\noindent {\textit{Locally stationary, globally nonstationary Gaussian process}}. {While the nonstationary approach of \cite{Risser2017} is not a Bayesian method, for comparison we include the convolution-based nonstationary model implemented in the \pkg{convoSPAT} package \citep{Risser2017}. This method obtains a globally nonstationary covariance function by estimating spatial dependence parameters locally and smoothing these local estimates using basis functions. We define a basis grid of 41 equally-spaced mixture component locations, and allow all aspects of the spatial dependence to vary over the domain (nugget, spatial variance, and anisotropy).}

\vskip2ex
\noindent {\textit{Approaches using novel methodology}}.
Using the methods implemented in \pkg{BayesNSGP}, we fit two statistical models to these data, both utilizing the sparse general Vecchia likelihood (\texttt{likelihood = "SGV"}) since the relatively large number of measurements makes using the exact likelihood infeasible. We fit both a stationary (anisotropic) and nonstationary covariance to these data (see Table \ref{modelComp}); for the nonstationary methods, we allow the spatial variance and anisotropy process to vary across CONUS while fixing the nugget variance to be a constant. In order to allow the signal-to-noise ratio to vary smoothly over the domain, we use the approximate GP model (with $K=50$ knots; see Figure \ref{app2_data}a) for the spatial variance process. From Section \ref{application1}, we expect the direction and magnitude of spatial dependence to depend on elevation, so we model the three anisotropy components as varying linearly with elevation (using componentwise regression). However, the elevation relationship may vary for, e.g., the Appalachian Mountains, relative to the Rocky Mountains or Cascades; therefore, we include longitude and the elevation/longitude interaction in the componentwise regression. 

For the MCMC, we specify a random walk sampler for the mean coefficient (stationary and nonstationary) and three block random walk samplers for the 4 coefficients (intercept, elevation, longitude, and the interaction) for each of the three anisotropy sub-processes. Regarding sampling of the latent GP effects for the $\sigma(\cdot)$ process in NGP-SGV: from the previous application, the best sampling scheme involved splitting the total number of knots into spatial sub-blocks of either 8 or 16 with random walk samplers on the hyperparameters. Similarly, here we break up sampling of the latent GP effects into five sub-blocks of nine to twelve parameters each (the location of the knots and their sampling groups are shown in Figure \ref{app2_data}a), with random walk samplers on the hyperparameters. 

\subsubsection{Computational time and efficiency}

\begin{table}[!t]
\caption{Details on the MCMC for each of the models fit to the precipitation rate data. The time is given in minutes per 1,000 posterior samples; ``\# Par" is the number of covariance parameters; efficiency is the number of effective samples per 100 seconds. Note: the LSGN time corresponds to the total time required to fit the model; posterior samples are not available for TGP because of the variable dimension of the parameter space. All times correspond to running the analysis on one core of a 12-core (Intel Xeon CPU E5520) machine with 128 GB memory.}
\begin{center}
\begin{tabular}{|c|c|c||c|c|c|}
\hline
  & & & \multicolumn{3}{|c|}{\textbf{Efficiency} [scalar or (min, max)]}  \\ \hline
\textbf{Model} & \textbf{Time} & \textbf{\# Par} & \textbf{$\tau$ } & \textbf{$\sigma$ } & \textbf{$\Sigma$ } \\
\hline \hline
TGP & 11.6   & n/a & n/a & n/a & n/a         \\ \hline
IGP & 11.4   & 3 & 2.28 & 0.10  & 0.09  \\ \hline
PP    & 2.4   & 3   & 6.23 & 3.59 & 3.62    \\ \hline
LSGN & 17.5$^*$ & 205   & -- & --   & --  \\ \hline 
AGP-SGV & 14.0  & 5 & 3.25  & 3.03 & 2.12 (1.63, 2.34)   \\ \hline
NGP-SGV & 42.7  & 66  & 0.69 & (0.003, 0.09) & (0.03, 0.30)   \\ \hline
\end{tabular}
\end{center}
\label{app2_time}
\end{table}%

As in Section \ref{application1}, we evaluate both the computational time and efficiency of the various MCMC schemes used to fit each model; see Table \ref{app2_time}. Again, the efficiency is defined as the number of effective samples from the posterior divided by the computational time needed to generate the samples. Here, the computational time is based on the time needed to generate all 40,000 samples for the Bayesian methods, and effective sample size is calculated based on the 10,000 post burn-in samples. Unfortunately, the number of parameters and hence efficiency are not available for TGP since the dimension of the parameter space is not constant over the MCMC.

In spite of having a very large number of parameters, the NGP-SGV model only takes approximately four times longer than IGP and PP.
Surprisingly, the computational time for IGP is similar to that of AGP-SGV, in spite of the fact that IGP uses the exact GP likelihood; the predictive process is very fast to fit, which yields high efficiencies. While the efficiency of the NGP-SGV is much smaller, this is likely mostly a function of the dimension of the parameter space. Furthermore, the efficiency of NGP-SGV for the anisotropy parameters is comparable with IGP.

\subsubsection{Cross-validation results}

\begin{table}[!t]
\caption{Mean square prediction error (MSPE) and continuous rank probability score (CRPS) for 10-fold cross-validation (see Table \ref{modelComp} for model label definitions); smaller scores indicate a better fit. The average score across all holdout sets is given, as well as the standard deviation of the scores, and the best scores are in bold.}
\begin{center}
\begin{tabular}{|p{2.5cm}||p{4.5cm}|p{4.5cm}|}
\hline
\textbf{Model} 	& \textbf{MSPE} 			& \textbf{CRPS} \\ \hline\hline 
TGP	        & 0.0552 (0.0091)   & 0.1073$^*$ (0.0059$^*$)              \\ \hline 
IGP	        & 0.0415 (0.0074)   & 0.1044 (0.0064)       \\ \hline 
PP	        & 0.0737 (0.0164)   & 0.1490 (0.0099)       \\ \hline 
LSGN	    & 0.0427 (0.0078)   & 0.0959$^*$ (0.0055$^*$)   \\ \hline 
AGP-SGV	    & 0.0406 (0.0066)   & 0.1023 (0.0063)       \\ \hline
NGP-SGV	    & \textbf{0.0351 (0.0060)}   & \textbf{0.0907 (0.0048)}            \\ \hline
\end{tabular}
\end{center}
\label{cvComp}
\begin{flushleft}
\vskip-2ex
{\scriptsize $^*$Assumes a Gaussian predictive distribution.}\\
\end{flushleft}
\end{table}%

Next, we compare the various statistical models in Table \ref{modelComp} using 10-fold cross validation (see Table \ref{cvComp}), where we randomly split the weather stations into 10 holdout sets. For each holdout set, we fit the various statistical models using measurements from the other nine holdout sets and generate posterior predictions for the held out locations. Then, we calculate the mean square prediction error (MSPE; using the posterior predictive mean, or kriging mean for LSGN) and the continuous rank probability score \citep[CRPS;][]{PropScoring} for each holdout location, and then average over all holdout locations. Finally, we show the mean score across holdout sets as well as the standard deviation of the ten scores; for both MSPE and CRPS, smaller scores indicate a better model fit. \cite{Kruger2016} outline several methods for estimating the CRPS for individual predictions based on the output from MCMC algorithms; for most of the models, we use the \pkg{scoringRules} package for \proglang{R} \cite[][version 0.9]{scoringRules} with the empirical CDF method to calculate the CRPS \citep{Kruger2016}. However, the \pkg{tgp} package does not provide posterior predictive samples as part of their output; thus, for TGP and LSGN (which is a non-Bayesian model) we calculate the CRPS assuming a Gaussian predictive density \citep[see][]{PropScoring} with mean and standard deviation corresponding to the posterior predictive mean and standard deviation (for TGP) or the kriging means and standard errors (for LSGN).

Under both scores, the nonstationary Gaussian process model with SGV likelihood (NGP-SGV) is identified as providing a best fit to the data, as it yields the smallest score and also has the smallest variability across scores. NGP-SGV even outperforms the highly flexible LSGN, indicating the importance of using covariates to characterize variability in the anisotropy process. The predictive process model performs the worst under both metrics, which is not overly surprising since this method fails to capture fine-scale spatial variability. Interestingly, the stationary spatial models IGP and AGP-SGV outperform the nonstationary TGP model; furthermore, AGP-SGV (which is anisotropic) provides a slight improvement over IGP (which is isotropic).

\subsubsection{Summaries of nonstationarity for NGP-SGV}

\begin{figure}[!t]
\begin{center}
\includegraphics[trim={0 0 0 0mm}, clip, width = 0.91\textwidth]{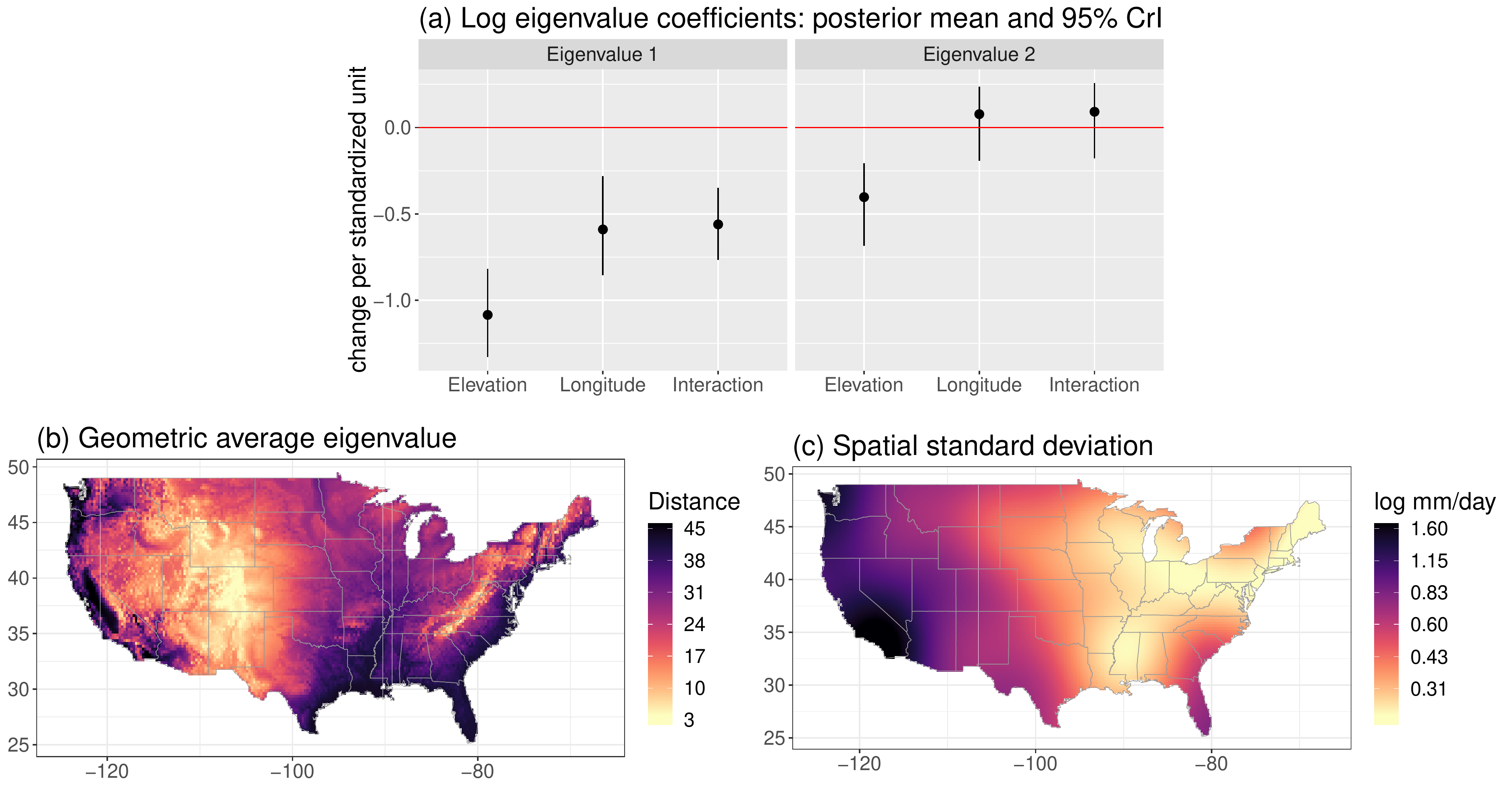}
\caption{Posterior mean and 95\% Bayesian credible intervals for the log eigenvalue coefficients (panel a), with a spatial map of the geometric average eigenvalue ($\sqrt{\lambda_1(\bfs)\lambda_2(\bfs)}$) (panel b). Panel c shows the posterior mean of the spatial standard deviation $\sigma(\cdot)$.}
\label{figure_app2summ}
\end{center}
\end{figure}

As a brief aside, an important benefit of our methodology is being able to summarize how and why the precipitation rate data exhibit nonstationarities across CONUS. For example, consider Figure \ref{figure_app2summ}, which shows posterior summaries for the anisotropy process $\bfSig(\cdot)$ and spatial standard deviation process $\sigma(\cdot)$. First, panel (a) shows the posterior mean and 95\% Bayesian credible interval for the regression coefficients on the log anisotropy eigenvalues (eigenvalues 1 and 2 refer to the first and secoond principal axes). Note that the coefficients for elevation in both axes are negative and significantly less than zero: this indicates that higher elevations have shorter length scales of spatial dependence. Interestingly, the interaction coefficient is only significantly different from zero in the first principal axis, which indicates that the relationship between elevation and spatial dependence does in fact change (at least for the first axis) across the CONUS. As another way to visualize the spatial variability in the anisotropy eigenvalues, a map of the geometric mean eigenvalue ($\sqrt{\lambda_1(\bfs)\lambda_2(\bfs)}$) is shown in Figure \ref{figure_app2summ}b. We can again see that the shortest length scales of dependence are in the highest elevations in the Rocky Mountains, and that length scales are much longer along the coasts.

Next, consider the posterior mean of the spatial standard deviation process $\sigma(\cdot)$ shown in Figure \ref{figure_app2summ}c. This map nicely illustrates the ``nonparametric'' regression framework that results from using a latent Gaussian process to characterize spatial variability in $\sigma(\cdot)$. However, also note that with only 50 knots, we can only characterize very large-scale variability in the spatial variance. Regardless, from the map we can see that the spatial variability is largest along the western coast of CONUS, with much less variability over the Ohio/Mississippi River Valleys and northeast US.

\begin{figure}[!t]
\begin{center}
\includegraphics[trim={0 0 0 0mm}, clip, width = 0.9\textwidth]{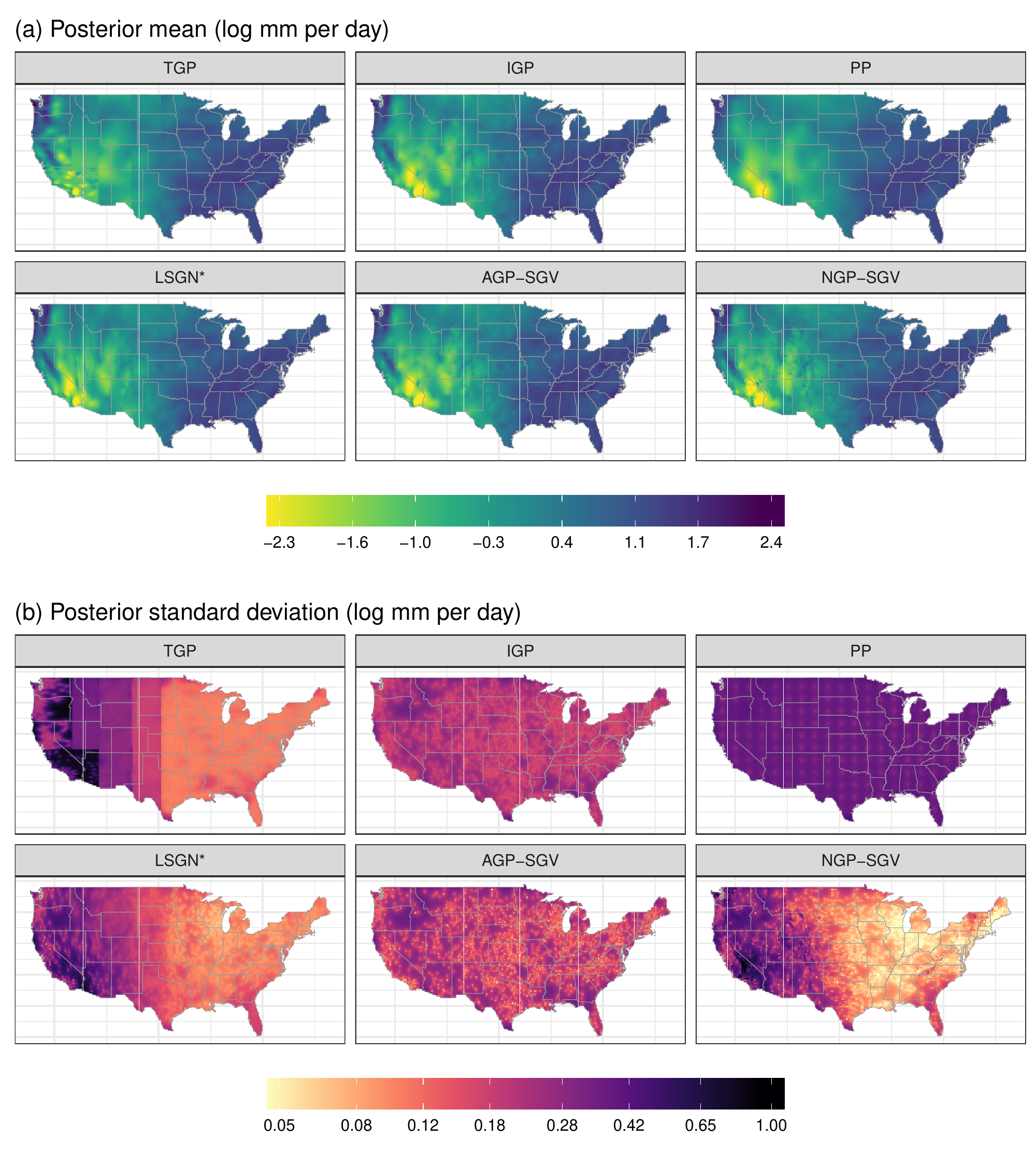}
\caption{Posterior mean and standard deviation (log mm day$^{-1}$) of the posterior predictive samples for total precipitation over CONUS in the 2018 water year, comparing the various models fit to these data. (Note: $^*$ indicates a Frequentist implementation.)}
\label{figure_app2pred}
\end{center}
\end{figure}

\subsubsection{Posterior prediction maps}

Finally, we explore a qualitative comparison of the various posterior predictive mean and standard deviation maps for each model, shown in Figure \ref{figure_app2pred}. Aside from TGP and PP, the posterior predictive mean maps look very similar for all models, with very low PR in southern California up through the Rocky Mountain states and much heavier PR in the Pacific Northwest, Sierra Nevada, and the eastern third of CONUS. However, there are clear differences in the posterior predictive standard deviations: standard errors for the stationary models IGP and AGP-SGV are driven by the location of the weather stations, and the PP standard errors show artifacts due to the position of the knot locations. Both NGP-SGV and LSGN much more flexibly capture spatial variability in the prediction errors, with very small uncertainty in the eastern United States and much larger uncertainty in the western United States; however, note that NGP-SGV has smaller uncertainty in the east, relative to LSGN. The TGP uncertainty captures a similar pattern (smaller uncertainty in the east and larger in the west), but these errors display artificial sharp boundaries due to the tree partitioning. These sharp boundaries likely contribute to the fact that stationary models IGP and AGP-SGV outperform TGP in the cross-validation.

\section{Discussion} \label{sec_discussion}

In this paper, we have developed a novel and highly flexible nonstationary covariance function along with a framework for incorporating the covariance function into approximate GP methods that enable fully Bayesian inference for high-dimensional data sets. The covariance function allows spatially-varying parameters to be specified in a variety of ways, including both regression-based and stochastically. Fully Bayesian inference along with posterior prediction are included in the corresponding \pkg{BayesNSGP} package, and we have demonstrated the features and computational properties of our methods using two data sets of varying size, from small (several hundred) to moderately large (several thousand). However, we note that we have used our software package with the NNGP likelihood to analyze a much larger data set with more than $50,000$ locations \citep[see][]{risser2020nonstationary} with only the computational resources of a personal laptop.

\begin{figure}[!t]
\begin{center}
\includegraphics[trim={0 0 0 0mm}, clip, width = 0.85\textwidth]{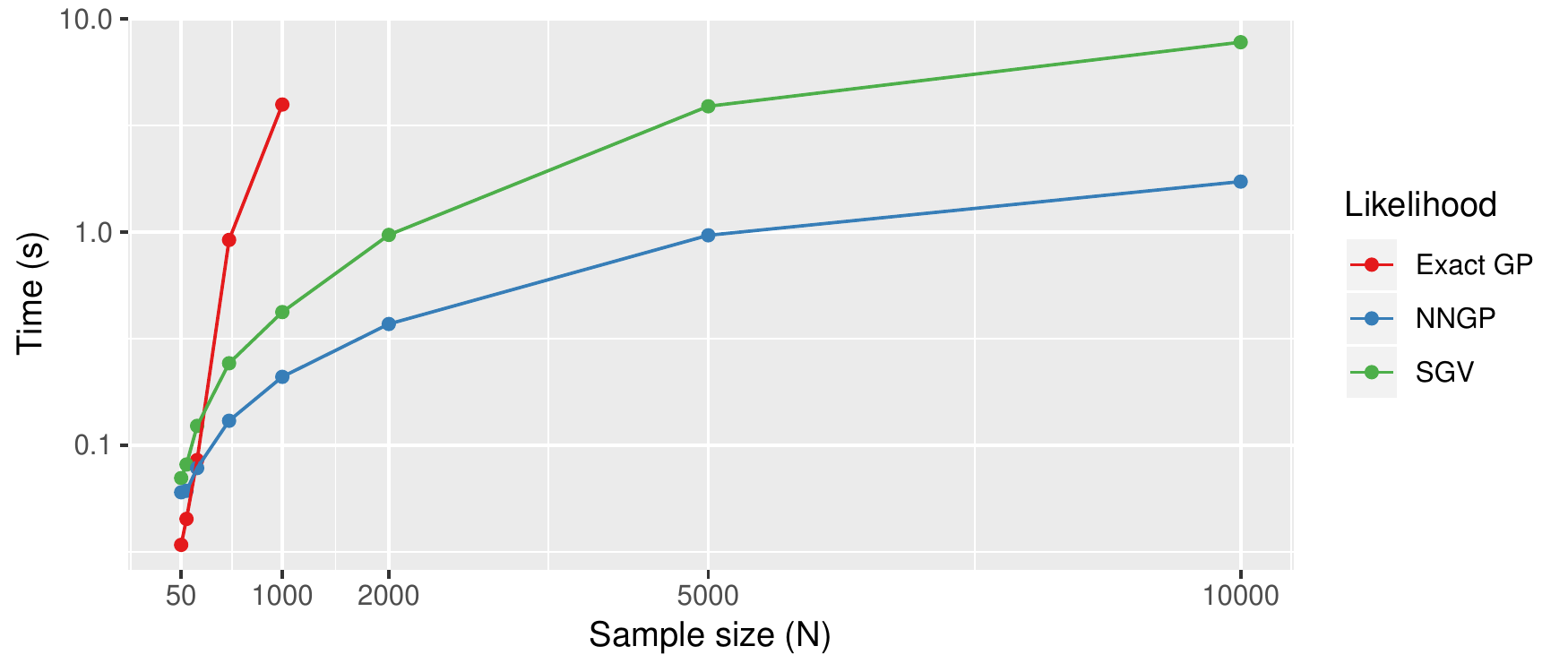}
\caption{Computational time for a single evaluation of the likelihood for a toy nonstationary model, for each likelihood method and a variety of sample sizes. Times correspond to a 1.6GHz Intel Core i5 machine with 16GB memory.}
\label{figure_compTime}
\end{center}
\end{figure}

As described in Section \ref{section4}, the approximate Gaussian process methods implemented in this paper and the \pkg{BayesNSGP} package (NNGP and SGV) present a trade-off between the quality of the likelihood approximation (SGV is known to provide a better approximation) and the computational speed (NNGP is faster). The fact that NNGP is faster than SGV remains true in our implementation for nonstationary Gaussian processes; as an illustration consider Figure \ref{figure_compTime}, which shows computational times for a single likelihood evaluation for a toy nonstationary example (for more information, see the Supplemental Materials). Note that the exact GP likelihood is actually the fastest for small sample sizes (up to about $N = 200$), while the NNGP is always faster than SGV even for very small data sets; the differences are a significant order of magnitude for $N=10,000$.

While our methodology enables approximate Gaussian process inference for ``large'' data sets, it should be noted that the methodology as implemented here is not extendable to modern ``massive'' data sets with many millions of measurements (\citealp{finley2018efficient} analyze a data set with over 28 million measurements, while \citealp{huang2019pushing} utilize high performance computing to conduct likelihood inference for more than 45 million measurements). For data sets of this size, there are non-negligible memory considerations for loading and manipulating the data; these issues are certainly still a barrier when using the \pkg{nimble} functionality. Nonetheless, we argue that our methods and correponding software package are still a novel and important contribution to the collection of available software for modeling (at least) moderately large data sets without requiring a customized computing environment or high performance computing: all of the analyses in this paper can be implemented off-the-shelf on a personal laptop.

\section*{Acknowledgements}

The authors would like to thank Dr. Chris Paciorek for helpful discussion and comments in the development of the methods and corresponding software package, and Dr. Dorit Hammerling for helpful feedback on development of the manuscript.

This work was supported by the Regional and Global Model Analysis Program of the Office of Biological and Environmental Research in the Department of Energy Office of Science under contract number DE-AC02-05CH11231. This document was prepared as an account of work sponsored by the United States Government. While this document is believed to contain correct information, neither the United States Government nor any agency thereof, nor the Regents of the University of California, nor any of their employees, makes any warranty, express or implied, or assumes any legal responsibility for the accuracy, completeness, or usefulness of any information, apparatus, product, or process disclosed, or represents that its use would not infringe privately owned rights. Reference herein to any specific commercial product, process, or service by its trade name, trademark, manufacturer, or otherwise, does not necessarily constitute or imply its endorsement, recommendation, or favoring by the United States Government or any agency thereof, or the Regents of the University of California. The views and opinions of authors expressed herein do not necessarily state or reflect those of the United States Government or any agency thereof or the Regents of the University of California.

\bibliographystyle{tfnlm}
\bibliography{BayesNSGP}

\clearpage
\appendix
\section{Supplemental figures and tables}
\label{appSupp}

\begin{table}[h]
\caption{MCMC sampling schemes used to draw from the posterior of the \cite{Paciorek2006} analysis described in Section \ref{app1Orig}. Block size refers to the number of latent anisotropy parameters sampled; joint/sep refers to whether the latent anisotropy parameters $\{w_1(\cdot), w_2(\cdot), w_3(\cdot)\}$ are sampled jointly or separately; RW/slice refers to whether the latent anisotropy hyperparameters $\{ \mu_\Sigma^k, \sigma_\Sigma^k, \phi_\Sigma^k: k = 1, 2\}$ are sampled using a random walk or slice sampler. The time is the number of hours required to run 100,000 MCMC iterations, and the minimum efficiency (``Min Eff'') is the effective sample size (out of the 5,000 thinned post burn-in saved samples) per hour of run time needed to generate all 100,000 MCMC samples. All times correspond to running the analysis on one core of a 12-core (Intel Xeon CPU E5520) machine with 128 GB memory.}
\begin{center}
\begin{tabular}{|c||c|c|c|c|c|} \hline
\textbf{Scheme}	& \textbf{Block size} & \textbf{Joint/Sep} & \textbf{RW/slice} & \textbf{Time (hr)} & \textbf{Min Eff}		\\ \hline\hline 
1 & 4 & Separate & RW & 47.60 & 0.06\\
\hline
2 & 8 & Separate & RW & 27.12 & 0.33\\
\hline
3 & 16 & Separate & RW & 16.50 & 0.49\\
\hline
4 & 32 & Separate & RW & 11.22 & 0.45\\
\hline
5 & 12 & Joint & RW & 20.15 & 0.39\\
\hline
6 & 24 & Joint & RW & 12.79 & 0.64\\
\hline
7$^\dagger$ & 48 & Joint & RW & 9.35 & \sout{0.85}\\
\hline
8$^\dagger$ & 96 & Joint & RW & 7.23 & \sout{1.41}\\
\hline
9 & 4 & Separate & Slice & 71.61 & 0.06\\
\hline
10 & 8 & Separate & Slice & 51.68 & 0.09\\
\hline
11 & 16 & Separate & Slice & 39.86 & 0.15\\
\hline
12 & 32 & Separate & Slice & 33.25 & 0.28\\
\hline
13 & 12 & Joint & Slice & 43.60 & 0.29\\
\hline
14 & 24 & Joint & Slice & 36.34 & 0.37\\
\hline
15$^\dagger$ & 48 & Joint & Slice & 32.62 & \sout{0.18}\\
\hline
16$^\dagger$ & 96 & Joint & Slice & 29.31 & \sout{0.51}\\
\hline
\end{tabular}
\end{center}
\label{app1_time}
\begin{flushleft}
\vskip-2ex
{\scriptsize $^\dagger$These samplers are excluded based on poor mixing of the MCMC.}\\
\end{flushleft}
\end{table}%

\begin{figure}[h]
\begin{center}
\includegraphics[trim={0 0 0 0mm}, clip, width = 0.8\textwidth]{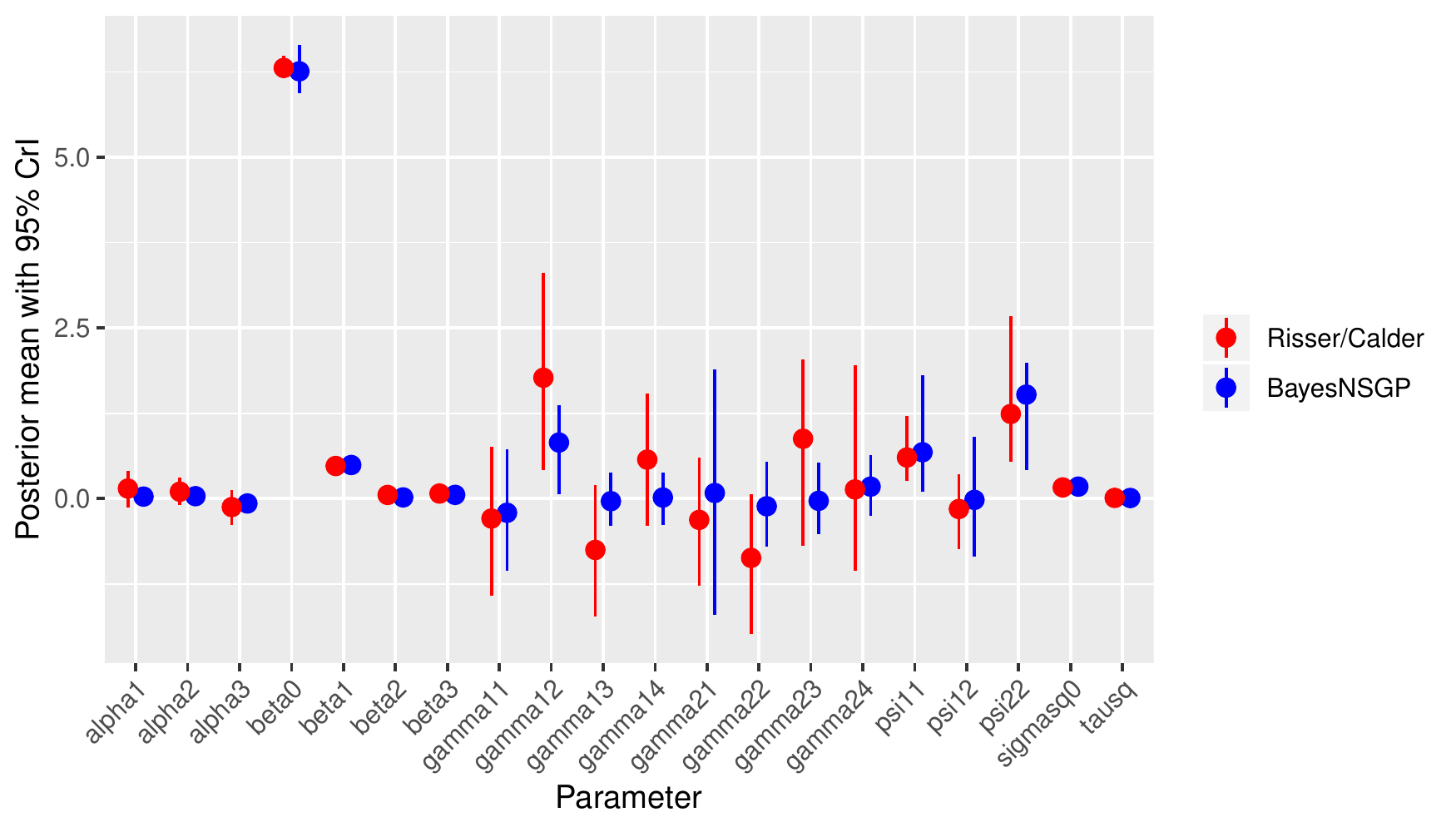}
\caption{Posterior mean and 95\% credible intervals for all statistical parameters in the covariance regression analysis of \cite{Risser2015}, comparing the original analysis as well as the \pkg{BayesNSGP} analysis provided in this paper.}
\label{compareEnv}
\end{center}
\end{figure}

\begin{figure}[!t]
\begin{center}
\includegraphics[trim={0 0 0 0mm}, clip, width = \textwidth]{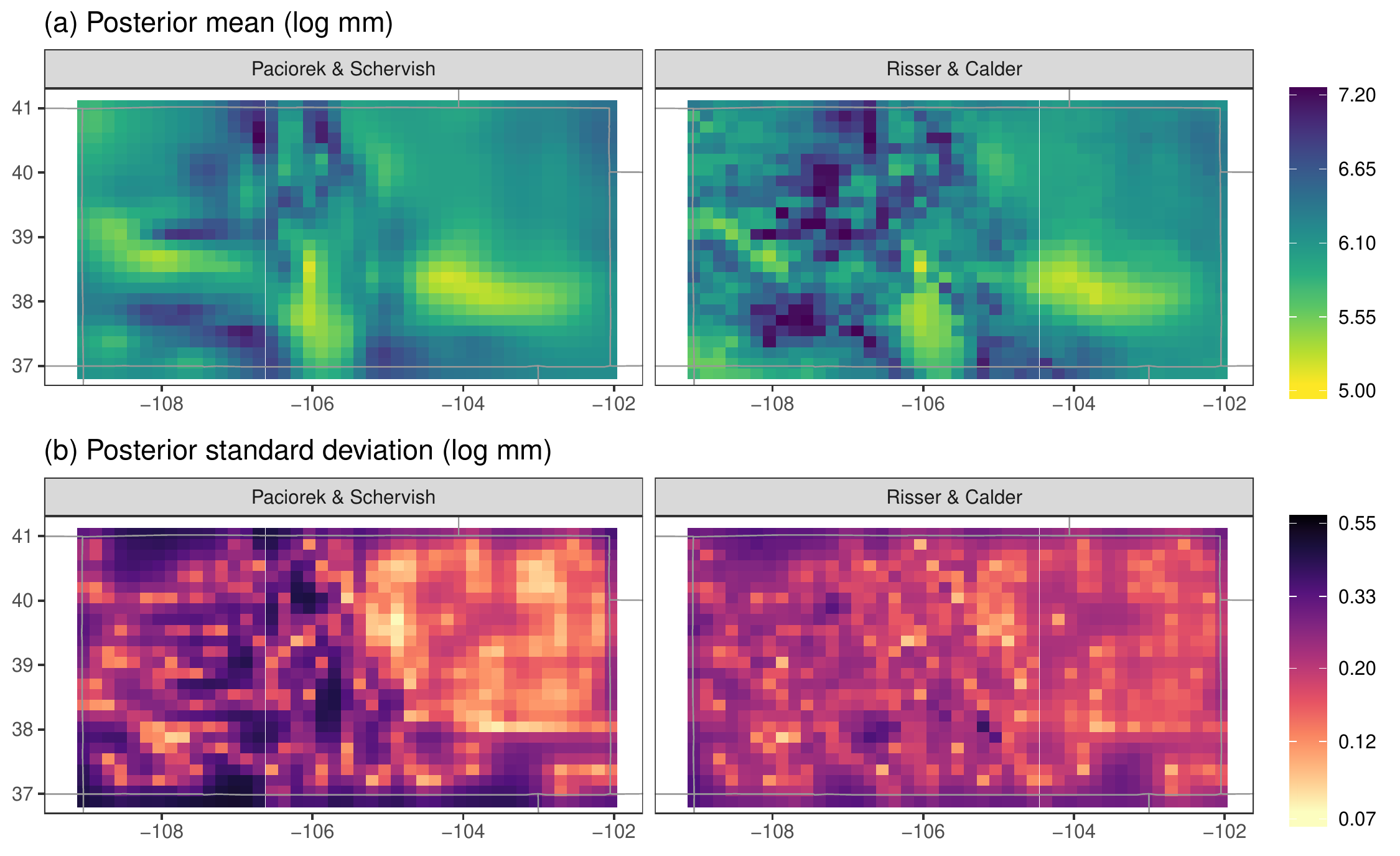}
\caption{Posterior mean and standard deviation (log mm) of the posterior predictive samples for both statistical models.}
\label{app1_postPred}
\end{center}
\end{figure}

\end{document}